\documentclass[aps,prl,onecolumn,showpacs,preprintnumbers]{revtex4}
\usepackage{amsmath}
\usepackage{dcolumn}
\usepackage{bm}
\usepackage{subfigure}
\usepackage{amsfonts}
\usepackage{amssymb}
\usepackage{makeidx}
\usepackage{epsfig}
\usepackage{graphicx}

\begin{document}

\title[]{Theory of non-local point transformations - Part 2: General form and
Gedanken experiment}
\author{Massimo Tessarotto}
\affiliation{Department of Mathematics and Geosciences, University of Trieste, Italy}
\affiliation{Institute of Physics, Faculty of Philosophy and Science, Silesian University
in Opava, Bezru\v{c}ovo n\'{a}m.13, CZ-74601 Opava, Czech Republic}
\author{Claudio Cremaschini}
\affiliation{Institute of Physics, Faculty of Philosophy and Science, Silesian University
in Opava, Bezru\v{c}ovo n\'{a}m.13, CZ-74601 Opava, Czech Republic}
\date{\today }

\begin{abstract}
The problem is posed of further extending the axiomatic construction
proposed in Part 1 for non-local point transformations mapping in each other
different curved space times. The new transformations apply to curved space
times when expressed in arbitrary coordinate systems. It is shown that the
solution permits to achieve an ideal (Gedanken) experiment realizing a
suitable kind of phase-space transformation on point-particle classical
dynamical systems. Applications of the theory are discussed both for
diagonal and non-diagonal metric tensors.
\end{abstract}

\pacs{02.40.Hw, 04.20.-q, 04.20.Cv}
\keywords{}
\maketitle





\section{1 - Introduction}

Following Ref.\cite{Part 1} (hereon referred to as "Part 1"), in this paper
further aspects are investigated concerning the extension of the functional
setting which lays at the\ basis of the standard formulation of General
Relativity (SF-GR), and Special Relativity (SR) as well \cite%
{Einstein1915,ein-1907,ein-1911,ein28,Einst,Cremaschini2015}. More
precisely, the issue is about the most general prescription of the class of
non-local point transformations (\emph{NLPT}) and related extended
GR-reference frames (\emph{extended GR-frames}) to be established between
two, in principle arbitrary, curved space-times.\textbf{\ }This will be
referred to here as \emph{general NLPT-theory}, in contrast to the \emph{%
special NLPT-theory}\textbf{\ }earlier developed in Ref.\cite{Part 1}.

In previous literature these transformation were identified with\emph{\
local point transformations} (LPT) and consequently necessarily mapping a
single space-time in itself only \cite%
{Einstein1915,Landau,Wheeler,wald,Synge}.\textbf{\ }Such a feature, which is
actually at the basis of SF-GR, i.e., Einstein's theory of gravitation,\ is
also of paramount importance in all relativistic theories, ranging from
classical to quantum electrodynamics, mechanics and theory of fields.
Nevertheless, in certain physical problems such as the \emph{Einstein's
Teleparallel approach to GR }(or \emph{TT-problem}), the introduction of a
new type of coordinate transformations, identified with the same NLPT
indicated above, is found to be mandatory. This refers to the of the theory
originally formulated by Einstein in 1928 \cite{ein28} in order to establish
a map between a generic connected and time-oriented curved space-time\textbf{%
\ }$(\mathbf{Q}^{4},g)$ and the flat time-oriented Minkowski space time $(%
\mathbf{Q}^{\prime 4},g^{\prime })\equiv (\mathbf{M}^{4},\eta )$ represented
in terms of orthogonal Cartesian coordinates (see Eqs.(\ref{CARTESIAN
COORDINATES-2}) below).\textbf{\ }In such a case its metric tensor is
identified with the corresponding Minkowski metric tensor $g_{\alpha \beta
}^{\prime }\equiv \eta _{\alpha \beta }=$diag$\left\{ 1,-1,-1,-1\right\} $.

As shown in Part 1, in particular, this means that it should always be
possible to represent such transformations\ in terms of real variables (see
for example Refs.\cite{NJ3,bambi,bambi2}).\textbf{\ } Also from Part 1 it
follows that the solution of the TT-problem involves in particular the
following two fundamental notions:

\emph{Notion \#1: the extended GR-frame.- }Departing from the customary
notion of GR-reference frame (or briefly GR-frame) traditionally adopted in
SF-GR, i.e., of a $4-$dimensional real curvilinear coordinate system $r^{\mu
}$\ to be established on $4-$dimensional Lorentzian space-times $(\mathbf{Q}%
^{4},g)$, the concept of \emph{extended GR-frame} is introduced. In each of
the two space-times $(\mathbf{Q}^{4},g)$ and\textbf{\ }$(\mathbf{Q}^{\prime
4},g^{\prime })$ this corresponds to identify such a notion with the
smoothly $s-$dependent phase-space state vectors%
\begin{eqnarray}
\mathbf{x}(s) &\equiv &\left\{ r^{\mu }(s),u^{\mu }(s)\equiv \frac{d}{ds}%
r^{\mu }(s)\right\} ,  \label{state-vectors-1} \\
\mathbf{x}^{\prime }(s) &\equiv &\left\{ r^{\prime \mu }(s),u^{\prime \mu
}(s)\equiv \frac{d}{ds}r^{\prime \mu }(s)\right\} ,  \label{state-vectors-2}
\end{eqnarray}%
which are defined at the same prescribed proper time $s$. Here $s$ by
assumption belongs to a suitable interval $I\subseteq
\mathbb{R}
$. Furthermore, $r^{\prime \mu }(s)\equiv r^{\prime \mu }\left\{ r(s),\left[
r,u\right] \right\} \in (\mathbf{Q}^{\prime 4},g^{\prime })$ and\textbf{\ }$%
r^{\mu }(s)\equiv r^{\mu }\left\{ r^{\prime }(s),\left[ r^{\prime
},u^{\prime }\right] \right\} \in (\mathbf{Q}^{4},g)$ identify, in terms of
in principle arbitrary coordinate systems, the corresponding $4-$positions
in the two space-times. Finally, $u^{\mu }(s)\equiv \frac{d}{ds}r^{\mu }(s)$
and $u^{\prime \mu }(s)\equiv \frac{d}{ds}r^{\prime \mu }(s)$ represent the
related $4-$velocities which span the corresponding tangent spaces.

\emph{Notion \#2: special NLPT-phase-space transformation -}\textbf{\emph{\ }%
}The second one is about the prescription of a suitable phase-space
transformation mapping in each other the two extended GR-frames $\left\{
r^{\mu }(s),u^{\mu }(s)\equiv \frac{d}{ds}r^{\mu }(s)\right\} $ and $\left\{
r^{\prime \mu }(s),u^{\prime \mu }(s)\equiv \frac{d}{ds}r^{\prime \mu
}(s)\right\} .$ For this purpose, in accordance with Einstein's TT-problem,
the curvilinear coordinates $r^{\mu }$ and $r^{\prime \mu }$ are
preliminarily identified with Cartesian coordinates, i.e., letting
respectively%
\begin{eqnarray}
r^{\mu } &\equiv &\left( (r^{0},\left( \mathbf{r}\equiv x,y,z\right) \right)
,  \label{CARTESIAN COORDINATES-1} \\
r^{\prime \mu } &\equiv &\left( (r^{0\prime },\left( \mathbf{r}^{\prime
}\equiv x^{\prime },y^{\prime },z^{\prime }\right) \right) .
\label{CARTESIAN COORDINATES-2}
\end{eqnarray}%
Then, the \emph{NLPT-phase-space transformation}\textbf{\emph{\ }}determined
in Part 1 is of the form%
\begin{equation}
\left\{
\begin{tabular}{l}
$\left\{ r^{\mu }(s),u^{\mu }(s)\right\} \rightarrow \left\{ r^{\prime \mu
}(s),u^{\prime \mu }(s)\right\} =\left\{ r^{\prime \mu }\left\{ r(s),\left[
r,u\right] \right\} ,\left( M^{-1}\right) _{\nu }^{\mu }(s)u^{\nu
}(s)\right\} ,$ \\
$\left\{ r^{\prime \mu }(s),u^{\prime \mu }(s)\right\} \rightarrow \left\{
r^{\mu }(s),u^{\mu }(s)\right\} =\left\{ r^{\mu }\left\{ r^{\prime }(s),%
\left[ r^{\prime },u^{\prime }\right] \right\} ,M_{\nu }^{\mu }(s)u^{\prime
\nu }(s)\right\} .$%
\end{tabular}%
\right.   \label{T-PHASE-2}
\end{equation}%
Here, again departing from SF-GR, the coordinate transformation $r^{\mu
}(s)\leftrightarrow r^{\prime \mu }(s)$ rather than being identified with a
local point transformation (LPT) acting on the same space-time $(\mathbf{Q}%
^{4},g)$, is realized by special NLPT mapping in each other two space-times%
\textbf{\ }$(\mathbf{Q}^{4},g)$ and $(\mathbf{Q}^{\prime 4},g^{\prime
})\equiv (\mathbf{M}^{\prime 4},\eta )$. In\ Lagrangian form these are of
the type%
\begin{equation}
\left\{
\begin{array}{c}
P_{S}\text{: }r^{\mu }(s)=r^{\prime \mu }(s_{o})+\int_{s_{o}}^{s}d\overline{s%
}M_{\nu }^{\mu }(s)u^{\prime \nu }(\overline{s}), \\
P_{S}^{-1}\text{: }r^{\prime \mu }(s)=r^{\mu }(s_{o})+\int_{s_{o}}^{s}d%
\overline{s}\left( M^{-1}\right) _{\nu }^{\mu }(s)u^{\nu }(\overline{s}),%
\end{array}%
\right.   \label{NLPT}
\end{equation}%
being\ the two space-times referred to \emph{the same coordinate systems }(%
\emph{Assumption }$\alpha $) and in Part 1 the latter were exclusively
identified with the Cartesian coordinates (\ref{CARTESIAN COORDINATES-1})-(%
\ref{CARTESIAN COORDINATES-2}). Here $M_{\nu }^{\mu }(s),\left(
M^{-1}\right) _{\nu }^{\mu }(s)$ denote the Jacobian matrix and its inverse,
both to be assumed of \emph{non-gradient type} (see related definitions in
Part 1). In the case in which the matrix $M_{\nu }^{\mu }$ (and $\left(
M^{-1}\right) _{\nu }^{\mu })$ are continuously connected to the identity $%
\delta _{\nu }^{\mu }$ this implies that%
\begin{equation}
\left\{
\begin{array}{c}
M_{\nu }^{\mu }=\delta _{\nu }^{\mu }+A_{\nu }^{\mu }(r^{\prime },r), \\
\left( M^{-1}\right) _{\nu }^{\mu }=\delta _{\nu }^{\mu }+B_{\nu }^{\mu
}(r,r^{\prime }),%
\end{array}%
\right.   \label{PPG-1}
\end{equation}%
with $\mathcal{A}_{\nu }^{\mu }$\textbf{\ }and\textbf{\ }$\mathcal{B}_{\nu
}^{\mu }$ being suitable transformation matrices. Hence, the special NLPT (%
\ref{NLPT}) yield the corresponding Lagrangian representations%
\begin{equation}
\left\{
\begin{array}{c}
r^{\mu }(s)=r^{\prime \mu }(s)+\int_{s_{o}}^{s}d\overline{s}A_{\nu }^{\mu
}(r^{\prime },r)u^{\prime \nu }(\overline{s}), \\
r^{\prime \mu }(s)=r^{\mu }(s)+\int_{s_{o}}^{s}d\overline{s}\mathcal{B}_{\nu
}^{\mu }(r^{\prime },r)u^{\nu }(\overline{s}).%
\end{array}%
\right.   \label{special-NLPT-1}
\end{equation}%
As discussed below (see Section 2), Eqs.(\ref{NLPT}), or equivalent (\ref%
{special-NLPT-1}), identify a group of transformations, denoted as \emph{%
special NLPT-group }$\left\{ P_{S}\right\} $\emph{,} established between $(%
\mathbf{M}^{4},\eta )$ and an in principle arbitrary curved space-time $(%
\mathbf{Q}^{4},g)$ in validity of Assumption $\alpha $.

In this paper we intend to investigate the physical bases for the
construction of more general non-local transformations, extending the class
of special NLPT prescribed by Eqs.(\ref{NLPT}) and holding in validity of
Assumption $\alpha $.\ The new transformations, denoted as\textbf{\ }\emph{%
general NLTP}\textbf{\ }and identifying the \emph{general NLPT-group }$%
\left\{ P_{g}\right\} $, will be achieved by introducing a suitable
axiomatic approach.\textbf{\ }These transformations will be permitted to map
two arbitrary distinct curved space-times $(\mathbf{Q}^{4},g)$ and\textbf{\ }%
$(\mathbf{Q}^{\prime 4},g^{\prime })$, each one to be represented in terms
of arbitrary coordinate systems, in particular generally different from the
Cartesian coordinate systems (\ref{CARTESIAN COORDINATES-1})-(\ref{CARTESIAN
COORDINATES-2}).

The fundamental issue which naturally arises in this connection is, of
course, whether these transformations may have a physical interpretation at
all. This would require in particular the identification of suitable
observable, i.e., classically measurable, dynamical variables. To answer
this question in a satisfactory and (hopefully) exhaustive way here we have
endeavoured to develop two partially independent routes.

\begin{itemize}
\item \emph{First route -} The first one is the search of a suitable \emph{%
Gedanken experiment }(GDE), namely an ideal measurement experiment, to
explicitly construct a general NLPT. As we intend to show, in fact, the same
GDE will permit: 1) The identification of the observable dynamical
variables, to be identified with the extended GR-frames (\ref%
{state-vectors-1}) and (\ref{state-vectors-2}) belonging respectively to the
curved space-times $(\mathbf{Q}^{4},g)$ and\textbf{\ }$(\mathbf{Q}^{\prime
4},g^{\prime })$. \ 2) The conceptual realization, and hence physical
interpretation, of an arbitrary transformation of the group of $\left\{
P_{g}\right\} $.

\item \emph{Second route -} The second route\ followed here, in order to
corroborate the GDE-based physical interpretation, is founded on the
development of selected applications of the general NLPT-theory, with
particular reference to the well-known theoretical issue related to the
diagonalization metric tensors associated with curved space-times. In the
literature such a problem is usally treated adopting the so-called
Newman-Janis algorithm to diagonalize non-diagonal. Such an algorithm is
frequently used in the literature for the purpose of investigating a variety
of standard or non-standard GR black-hole solutions (Bambi et al., 2013;
Toshmatov et al., 2014; Modesto et al., 2010). These include a number of
problems which have remained unsolved to date and appears again of critical
importance in GR. In particular:
\end{itemize}

\begin{enumerate}
\item \emph{Problem \#P}$_{1}$ - First, the fact that the Newman-Janis
algorithm is complex, so that the transformed coordinates are complex too.
This inhibits their objective physical interpretation in terms of
observables.

\item \emph{Problem \#P}$_{2}$ - The fact that again the diagonalization
problem at the basis of the same transformation cannot be solved in the
framework of the validity of the LPT-GCP.

\item \emph{Problem \#P}$_{3}$ - The physical meaning of the transformation:
one cannot ignore that fact that there is no clear understanding regarding
its physical interpretation and ultimately as to why the algorithm should
actually work at all.

\item \emph{Problem \#P}$_{4}$ - Finally, despite the obvious fact that the
Teleparallel transformation provides in principle also a solution to the
diagonalization problem, there is no clear connection emerging between the
same transformation and the Newman-Janis algorithm.
\end{enumerate}

The goal of this paper is to address specifically \emph{Problems} \emph{\#P}$%
_{1}-$\emph{\#P}$_{4},$ a task which has remain essentially un-challaged to
date. These problems are investigated based on the adoption of a suitable
realization of non-local point transformations (NLPT) acting on appropriate
extended GR-frames which are defined with respect to prescribed space-times.
For such a purpose the determination is required of the group of \emph{%
general non-local point transformations} (general NLPT) connecting subsets
of two generic curved space-times $(\mathbf{Q}^{4},g)$ and $(\mathbf{Q}%
^{\prime 4},g^{\prime }).$\ The task posed here involves also their \emph{%
physical interpretation} based on a suitable Gedanken experiment. This
refers, in particular, to three distinct issues:

A) The possible conceptual realization of a measure experiment (Gedanken
experiment), simulating the action of a generic NLPT on a GR-reference frame
on the physical space-time.

B) The prescription of the family of NLPT, exclusively based on a suitable
set of mathematical, i.e., axiomatic, prescriptions, which should be
nevertheless physically realizable in principle for arbitrary GR-reference
frames which are defined with respect to a prescribed space-time.

C) As an illustration of the theory, the explicit construction of possible
physically-relevant transformations of the group $\left\{ P_{g}\right\} ,$
with special reference to the problem of diagonalization of non-diagonal
metric tensors associated with rotating black holes.

As we intend to show, both routes will ultimately enable us to demonstrate
the interpretation and physical consistency of the general NLPT-theory
developed here, the connection with the analogous formulation holding for
special NLPT (presented in Part 1) and - most important in our view - to
display the explicit construction method of non-local transformations which
mutually map in each other a variety of curved space-times.

\subsection{An example from SR}

Consider as a preliminary illustration of the issue the classical dynamical
system (CDS) describing the dynamics of single point-particles in the
special relativity (SR) setting, i.e., in the time-oriented Minkowski
space-time. A possible Gedanken experiment concerns the representation of
the same CDS performing a suitable reference-frame transformation. We shall
distinguish - in such a process - both the so-called active and passive
viewpoints of the transformation, i.e., in which either a point particle
evolves in time (\textquotedblleft moves\textquotedblright )\ or the
reference frame itself changes, respectively. In order to define properly
the two viewpoints let us introduce the\textbf{\ }$4-$displacement and
corresponding $4-$velocity transformation of the type:%
\begin{equation}
\left\{
\begin{tabular}{l}
$dr^{\mu }=\mathcal{J}_{\nu }^{\mu }dr^{\prime \nu },$ \\
$dd^{\prime \mu }=\left( \mathcal{J}^{-1}\right) _{\nu }^{\mu }dr^{\nu },$%
\end{tabular}%
\right.  \label{4-displacement transf}
\end{equation}%
\begin{equation}
\left\{
\begin{tabular}{l}
$u^{\mu }=\mathcal{J}_{\nu }^{\mu }u^{\prime \nu },$ \\
$u^{\prime \mu }=\left( \mathcal{J}^{-1}\right) _{\nu }^{\mu }u^{\nu }.$%
\end{tabular}%
\right.  \label{4-velocity transf}
\end{equation}%
Eqs.(\ref{4-velocity transf}) can be viewed as a Gedanken experiment (GDE)
advancing in time separately the states $\left\{ r^{\mu }(s),u^{\mu
}(s)\right\} $\textbf{\ }and\textbf{\ }$\left\{ r^{\prime \mu }(s),u^{\prime
\mu }(s)\right\} $.\ To elucidate this point consider the following two
CDS's:%
\begin{equation}
\left\{ r^{\mu }(s_{o}),u^{\mu }(s_{o})\right\} \leftrightarrow \left\{
r^{\mu }(s),u^{\mu }(s)\right\} ,  \label{CDS-0}
\end{equation}%
\begin{equation}
\left\{ r^{\prime \mu }(s_{o}),u^{\prime \mu }(s_{o})\right\}
\leftrightarrow \left\{ r^{\prime \mu }(s),u^{\prime \mu }(s)\right\} ,
\label{CDS-1}
\end{equation}%
which are assumed to be prescribed for all $s_{o},s\in I$. Assuming validity
of Eqs.(\ref{4-displacement transf}) and (\ref{4-velocity transf}) it
follows that the two states\textbf{\ }$\left\{ r^{\mu }(s),u^{\mu
}(s)\right\} $ and $\left\{ r^{\prime \mu }(s),u^{\prime \mu }(s)\right\} $%
\textbf{\ }are\ manifestly not independent. Indeed, the same CDS's are not
independent, as it follows at once by direct inspection of Eqs.(\ref%
{4-velocity transf}) and (\ref{4-displacement transf}). In particular, the
first one (\ref{CDS-0}) (and respectively the second one (\ref{CDS-1})) are
obtained by considering the state\textbf{\ }$\left\{ r^{\prime \mu
}(s),u^{\prime \mu }(s)\right\} $ (or correspondingly $\left\{ r^{\mu
}(s),u^{\mu }(s)\right\} $) as prescribed. As discussed at length in the
following Sections, the two choices will be referred to as the \emph{active}
and\emph{\ passive viewpoints }in which the GDE can be considered, more
precisely: A) In the \emph{active viewpoint}\textbf{\ }the state\textbf{\ }$%
\left\{ r^{\mu }(s),u^{\mu }(s)\right\} $ acting on the curved
("transformed") space-time evolves in time with $\left\{ r^{\prime \mu
}(s),u^{\prime \mu }(s)\right\} $, the "background" state defined in the
Minkowski space-time being considered a prescribed smooth $s-$function and
generating the phase-space flow. B) In the \emph{passive viewpoint} the state%
\textbf{\ }$\left\{ r^{\mu }(s),u^{\mu }(s)\right\} $ is considered a
prescribed smooth function of $s,$ so that the background state $\left\{
r^{\prime \mu }(s),u^{\prime \mu }(s)\right\} $ must evolve in time
accordingly.

For definiteness, let us consider for the Jacobian matrix $\mathcal{J}_{\nu
}^{\mu }$, with $\left( \mathcal{J}^{-1}\right) _{\nu }^{\mu }$ being its
inverse, a realization which corresponds to a boost transformation, i.e., a
space-time rotation for which $\mathcal{J}_{\nu }^{\mu }\equiv \mathcal{J}%
_{\nu }^{\mu }(s)$, where%
\begin{equation}
\mathcal{J}_{\nu }^{\mu }(s)=\left\vert
\begin{array}{cccc}
\gamma (s) & -\beta (s)\gamma (s) & 0 & 0 \\
-\beta (s)\gamma (s) & \gamma (s) & 0 & 0 \\
0 & 0 & 1 & 0 \\
0 & 0 & 0 & 1%
\end{array}%
\right\vert ,
\end{equation}%
while $\gamma (s)$ and $\beta (s)$ are the Lorentz and relativistic factors $%
\gamma (s)=1/\sqrt{1-\beta ^{2}(s)}$, $\beta (s)\equiv \left\vert \mathbf{v}%
(s)\right\vert /c$ and $\mathbf{v}(s)$ denotes the spatial components of a
local and non-uniform reference velocity. In particular, let us require that
$\mathbf{v}(s)$ is parametrized in terms of the arc length $s$, to be
established on a suitable time-like word-line $r^{\mu }$ (see below). It
follows that by construction $u^{\mu }$ and $u^{\prime \mu }$ belong to
different tangent spaces defined with respect to the same Minkowskian
space-time, since by construction the identity%
\begin{equation}
\eta _{\alpha \beta }\frac{dr^{\alpha }}{ds}\frac{dr^{\beta }}{ds}=\eta
_{\alpha \beta }\frac{dr^{\prime \nu }}{ds}\frac{dr^{\prime \mu }}{ds}
\end{equation}%
manifestly holds. The corresponding coordinate transformations $r^{\alpha
}(s)$\ $\rightarrow r^{\prime \alpha }(s)$ and its inverse, both defined in $%
(M^{4},\eta )$ and generated by integrating the $4-$velocity transformations
(\ref{4-velocity transf}) along arbitrary time-like world lines of $%
(M^{4},\eta )$, are manifestly of the type indicated above (see Eqs.(\ref%
{NLPT})) and therefore identify a particular possible realization of NLPT.
It follows that Eqs.(\ref{NLPT}) can be interpreted as performed as a result
of the said Gedanken experiment. More precisely: A) In the \emph{active
viewpoint} a point-particle endowed with a $4-$position $r^{\prime \nu }$\
(or $r^{\nu }$) acquires a displacement which carries it to the transformed $%
4-$position $r^{\mu }$\ (or $r^{\prime \mu }$\ respectively), by means of a
suitable dynamical flow of some kind producing also such a change in the
particle $4-$position.\ B) In the \emph{passive viewpoint} the
point-particle $4-$position remains invariant, while the reference frame
changes in such a way that the $4-$position $r^{\prime \nu }$\
(respectively, $r^{\nu }$) is transformed to $r^{\mu }$\ ($r^{\prime \mu }$).

This simple example further supports the earlier discussion reported in Part
1 regarding the asserted physical inadequacy of\ the traditional concept of
reference frame (the so-called GR-frame) adopted in particular in the
context of GR, i.e., of a coordinate system based on the $4-$position $%
r\equiv \left\{ r^{\mu }\right\} $ only, which is founded - in turn - on the
adoption of purely \textit{local} coordinate transformations. The rationale
behind the issue\textbf{\ }considered here lies on the Einstein equivalence
principle (EEP, \cite{Einst}) itself. This is actually realized by two
separate propositions, which in the form presently known must both be
ascribed to Albert Einstein's 1907 original formulation \cite{ein-1907} (see
also Ref.\cite{ein-1911}). In Einstein's original approach this actually is
realized by the following two distinct claims stating: a) the equivalence
between accelerating frames and the occurrence of gravitational fields (see
also Ref.\cite{Einst}); b) that \textquotedblleft local effects of motion in
a curved space (gravitation)\textquotedblright\ should be considered\ as\
\textquotedblleft \emph{indistinguishable from those of an accelerated
observer in flat space}\textquotedblright\ \cite{ein-1907,ein-1911}.

This motivates us to search, based on physical first principles, for a
development of the subject which eventually should/might permit one:

\begin{itemize}
\item To determine the most general representation for the phase-space
transformations connecting \emph{physical reference frames}, to be referred
to as \emph{general NLPT-phase-space transformations.}

\item To address the main related mathematical and physical implications.
For generality these will be investigated in the framework of GR, since the
latter by construction encompasses also SR.
\end{itemize}

Under such premises, and to better elucidate the scope and potential
physical relevance of the issue indicated above it must be noticed that the
present work belongs to the class of studies aimed at extending in the
context of GR and classical field theory the classical notions of local
dynamics and local field interactions, with the precise goal of including in
these theories various types of non-local phenomena.\ Recent literature
investigations in this category are several. We refer to Part 1 for further
discussions on the matter.

However, an instance worth to be mentioned and most relevance in the present
discussion concerns the Einstein teleparallelism\textbf{\ }\cite{ein28}). In
fact, as discussed in Part 1, the teleparallel problem lying at the basis of
such an approach cannot be solved in the framework of GCP and actually
requires the introduction of a new functional setting for GR
(NLPT-functional setting) based on the introduction of suitable NLPT.

As shown below, the new approach proposed here, based on the introduction of
suitable NLPTs,\textbf{\ }permits to cast light on non-local phenomena which
can occur in GR due to the choice of the GR-reference frames.

\subsection{Goals of the paper}

Given these premises, we are now in position to state in detail the
structure of the present manuscript, pointing out the goals posed in each of
the following sections which are accordingly listed below.

\begin{enumerate}
\item \emph{GOAL \#1 - }The first one, discussed in\textbf{\ }Section 2,
concerns the development of the theory of \emph{general NLPT} which permits
to map in each other in principle arbitrary space-times. The connection with
the special NLTP-theory earlier developed in Part 1\ is displayed.

\item \emph{GOAL \#2 - }In Section 3 a Gedanken experiment interpretation
and physical implications of the general NLPT-theory developed here are
proposed for the phase-space transformations generated by the group of
general NLPT.

\item \emph{GOAL \#3 - }In Section 4, the application is considered of the
theory of general NLPT to the mapping of diagonal metric tensor expressed in
arbitrary curvilinear coordinates.

\item \emph{GOAL \#4 - }In Section 5, the goal is posed of addressing the
diagonalization problem for non-diagonal metric tensors occurring in GR.

\item \emph{GOAL \#5 - }Finally, in Section 6 the main conclusions of the
paper are drawn.
\end{enumerate}

\section{2 - Theory of general n\textbf{on-local point transformations on
manifolds}}

In this section the following representation problem is posed for such a
theory: this lies in the search of\ the most general form which these point
transformations and their theory, earlier pointed out in Part 1, can take.
In the following these will be referred to as \emph{general NLTP} and \emph{%
general NLPT-theory} respectively.

More precisely, the new transformations should realize a mapping between two
arbitrary connected and time-oriented\textbf{\ }$4-$dimensional curved
space-times $(\mathbf{Q}^{4},g)$ and\textbf{\ }$(\mathbf{Q}^{\prime
4},g^{\prime })$ when they are referred to arbitrary curvilinear coordinate
systems. For this purpose we shall require that by construction the new
transformations between $(\mathbf{Q}^{4},g)$ and $(\mathbf{Q}^{\prime
4},g^{\prime })$ determine a suitably-prescribed real diffeomorphism of the
general form%
\begin{equation}
P_{g}:r^{\prime \mu }(s)\rightarrow r^{\mu }(s)=P_{g}^{\mu }\left( r^{\prime
},u^{\prime },\left[ r^{\prime },u^{\prime },\frac{D^{\prime }u^{\prime }}{Ds%
}\right] ,s\right) ,  \label{pg1}
\end{equation}%
with inverse transformation%
\begin{equation}
P_{g}^{-1}:r^{\mu }(s)\rightarrow r^{\prime \mu }(s)=P_{g}^{-1}{}^{\mu
}\left( r,u,\left[ r,u,\frac{Du}{Ds}\right] ,s\right) ,  \label{pg2}
\end{equation}%
the square brackets denoting appropriate non-local dependences. In
particular, here $r^{\prime }\equiv \left\{ r^{\prime \mu }\right\} ,r\equiv
\left\{ r^{\mu }\right\} ,u^{\prime }\equiv \left\{ u^{\prime \mu }\right\}
,u\equiv \left\{ u^{\mu }\right\} ,$ while $\frac{D^{\prime }u^{\prime }}{Ds}%
\equiv \frac{D^{\prime }u^{\prime \mu }}{Ds}$and $\frac{Du}{Ds}\equiv \frac{%
Du^{\mu }}{Ds}$ identify as usual the covariant derivatives defined in the
two space-times $(\mathbf{Q}^{\prime 4},g^{\prime })$ and\textbf{\ }$(%
\mathbf{Q}^{4},g)$ respectively.

For definiteness, we shall also assume that Eqs.(\ref{pg1}) and (\ref{pg2})
are also consistent with the requirement that the proper-time $s$ appearing
in both equations satisfies the Riemann distance condition%
\begin{equation}
ds^{2}=g_{\mu \nu }(r)dr^{\mu }dr^{\nu }=g_{\mu \nu }^{\prime }(r^{\prime
})dr^{\prime \mu }dr^{\prime \nu },  \label{Riemann distance}
\end{equation}%
both prescribed in terms of real and suitably-smooth functions of $s$ and
hence warranting also the mass-shell kinematic constraints
\begin{equation}
\begin{array}{c}
u^{\mu }u^{\nu }g_{\mu \nu }(r)=1, \\
u^{\prime \mu }u^{\prime \nu }g_{\mu \nu }^{\prime }(r^{\prime })=1.%
\end{array}
\label{mass-shell conditions}
\end{equation}%
It is immediate to notice that an obvious possible realization of the
transformations (\ref{pg1}) and (\ref{pg2}) is obtained simply by
considering explicitly $s-$dependent smooth real transformations of the type%
\begin{eqnarray}
P_{g} &:&r^{\mu }\rightarrow r^{\prime \mu }=r^{\prime \mu }(r,s),
\label{T-1g} \\
P_{g}^{-1} &:&r^{\prime \mu }\rightarrow r^{\mu }=r^{\mu }(r^{\prime },s),
\label{T-2-g}
\end{eqnarray}%
defined for all $s\in I$.\textbf{\ }Again, for $dr^{\mu }$\ and $u^{\mu
}\equiv \frac{dr^{\mu }}{ds}$\ transformations of the type (\ref%
{4-displacement transf}) and (\ref{4-velocity transf}) are implied.\textbf{\
}However, the Jacobians are of the type $\mathcal{J}_{\nu }^{\mu }(r^{\prime
},s)$\ and $\left( \mathcal{J}^{-1}\right) _{\nu }^{\mu }(r,s)$\ and read
respectively%
\begin{eqnarray}
\mathcal{J}_{\nu }^{\mu }(r^{\prime },s) &\equiv &\frac{\partial r^{\mu
}(r^{\prime },s)}{\partial r^{\prime \nu }}+\frac{\partial r^{\mu
}(r^{\prime },s)}{\partial s}g_{\alpha \nu }^{\prime }u^{\prime \alpha },
\label{Jac-1g} \\
\left( \mathcal{J}^{-1}\right) _{\nu }^{\mu }(r,s) &\equiv &\frac{\partial
r^{\prime \mu }(r,s)}{\partial r^{\nu }}+\frac{\partial r^{\prime \mu }(r,s)%
}{\partial s}g_{\alpha \nu }u^{\alpha },  \label{Jac-2g}
\end{eqnarray}%
thus loosing their gradient form (see Eqs.(4) and (5) in Part 1).\textbf{\ }%
Nevertheless, it is obvious that transformations of the type indicated above
generally imply the violation of\ the Riemann-distance constraint (\ref%
{Riemann distance}).

On the other hand, once the implications of the same equation are properly
taken into account the representation problem posed here can be readily
solved. Consider in fact again Eq.(\ref{Riemann distance}).\textbf{\ }Due to
the arbitrariness of $r\equiv \left\{ r^{\mu }\right\} $ as well of $s$ and $%
dr^{\mu }$ it follows that the same equation requires simultaneously that%
\begin{equation}
\left\{
\begin{tabular}{l}
$d^{\mu }=M_{(g)\nu }^{\mu }d^{\prime \nu },$ \\
$d^{\prime \mu }=\left( M_{(g)}^{-1}\right) _{\nu }^{\mu }d^{\nu },$%
\end{tabular}%
\right.  \label{implication-1}
\end{equation}%
and%
\begin{equation}
\left\{
\begin{tabular}{l}
$g_{\mu \nu }(r)=\left( M_{(g)}^{-1}\right) _{\mu }^{\alpha }\left(
M_{(g)}^{-1}\right) _{\nu }^{\beta }g_{\mu \nu }^{\prime }(r^{\prime }),$ \\
$g_{\mu \nu }^{\prime }(r^{\prime })=M_{(g)\mu }^{\alpha }M_{(g)\nu }^{\beta
}g_{\mu \nu }(r),$%
\end{tabular}%
\right.  \label{implication-2}
\end{equation}%
must hold, with $M_{(g)\nu }^{\mu }$ denoting a suitable and still
to-be-determined real Jacobian matrix and $\left( M_{(g)}^{-1}\right) _{\nu
}^{\mu }$ being its inverse. Therefore, Eqs.(\ref{implication-1}) imply that
Eqs.(\ref{T-1g}) and (\ref{T-2-g}) must recover the form (\ref{NLPT}), while
Eqs.(\ref{implication-2}) require that necessarily takes the form
\begin{equation}
\left\{
\begin{array}{c}
M_{(g)\nu }^{\mu }(s)=M_{(g)\nu }^{\mu }(r^{\prime }(s),r(s)), \\
\left( M_{(g)}^{-1}\right) _{\nu }^{\mu }(s)=\left( M_{(g)}^{-1}\right)
_{\nu }^{\mu }(r(s),r^{\prime }(s)),%
\end{array}%
\right.  \label{Jacobian-NLPT}
\end{equation}%
i.e., they can only be functions of $r^{\prime }(s)$ or respectively $r(s).$
More precisely, on the rhs of the first (second) equation $r(s)$ ($r^{\prime
}(s)$) must be considered as a function of $r^{\prime }(s)$ (respectively of
$r(s)$) determined by means of an equation analogous to that holding for
special NLTP, i.e., Eqs.(\ref{NLPT}), namely%
\begin{equation}
\left\{
\begin{array}{c}
P_{g}\text{: }r^{\mu }(s)=r^{\prime \mu }(s_{o})+\int_{s_{o}}^{s}d\overline{s%
}M_{(g)\nu }^{\mu }(s)u^{\prime \nu }(\overline{s}), \\
P_{g}^{-1}\text{: }r^{\prime \mu }(s)=r^{\mu }(s_{o})+\int_{s_{o}}^{s}d%
\overline{s}\left( M_{(g)}^{-1}\right) _{\nu }^{\mu }(s)u^{\nu }(\overline{s}%
).%
\end{array}%
\right.  \label{g-NLPT}
\end{equation}%
This will be referred to as \emph{general NLPT}. \ The corresponding
phase-space map analogous to Eq.(\ref{T-PHASE-2}), namely%
\begin{equation}
\left\{
\begin{tabular}{l}
$\left\{ r^{\mu }(s),u^{\mu }(s)\right\} \rightarrow \left\{ r^{\prime \mu
}(s),u^{\prime \mu }(s)\right\} =\left\{ r^{\prime \mu }\left\{ r(s),\left[
r,u\right] \right\} ,\left( M_{(g)}^{-1}\right) _{\nu }^{\mu }(s)u^{\nu
}(s)\right\} ,$ \\
$\left\{ r^{\prime \mu }(s),u^{\prime \mu }(s)\right\} \rightarrow \left\{
r^{\mu }(s),u^{\mu }(s)\right\} =\left\{ r^{\mu }\left\{ r^{\prime }(s),%
\left[ r^{\prime },u^{\prime }\right] \right\} ,M_{(g)\nu }^{\mu
}(s)u^{\prime \nu }(s)\right\} ,$%
\end{tabular}%
\right.  \label{general NLPT-phase-space-map}
\end{equation}%
will be denoted as \emph{general NLPT-phase-space transformation}. Then the
following result holds.

\bigskip

\textbf{THM.1 - Realization of the general NLPT-group}\emph{\ }$\left\{
P_{g}\right\} $\emph{.}

\emph{The group }$\left\{ P_{g}\right\} $\emph{\ of general NLPT of the type
Eqs.(\ref{1B1}) can always be realized by means of Jacobians }$M_{(g)\nu
}^{\mu }$ \emph{and} $\left( M_{(g)}^{-1}\right) _{\nu }^{\mu }$\emph{of the
form}%
\begin{equation}
\left\{
\begin{array}{c}
M_{(g)\nu }^{\mu }=\frac{\partial g_{A}^{\mu }(r^{\prime })}{\partial
r^{\prime \nu }}+A_{(g)\nu }^{\mu }(r^{\prime },r), \\
\left( M_{(g)}^{-1}\right) _{\nu }^{\mu }=\frac{\partial f_{A}^{\mu }(r)}{%
\partial r^{\nu }}+B_{(g)\nu }^{\mu }(r,r^{\prime }),%
\end{array}%
\right.  \label{g-PPG-1}
\end{equation}%
\emph{with} $A_{(g)\nu }^{\mu }(r^{\prime },r)$ \emph{and }$B_{(g)\nu }^{\mu
}(r,r^{\prime })$\emph{\ being suitable transformation matrices. As a
consequence, an arbitrary general NLPT can be represented as}%
\begin{equation}
\left\{
\begin{array}{c}
P_{g}\text{: }r^{\mu }(s)=g_{A}^{\mu }(r^{\prime }(s))+\int_{s_{o}}^{s}d%
\overline{s}A_{(g)\nu }^{\mu }(s)u^{\prime \nu }(\overline{s}), \\
P_{g}^{-1}\text{: }r^{\prime \mu }(s)=f_{A}^{\mu }(r(s))+\int_{s_{o}}^{s}d%
\overline{s}B_{(g)\nu }^{-1\mu }(s)u^{\nu }(\overline{s}).%
\end{array}%
\right.  \label{g-NLPT-2}
\end{equation}%
\emph{Proof - }In fact, given validity of Eqs.(\ref{g-PPG-1}) it follows for
example that%
\begin{equation}
r^{\mu }(s)=r^{\prime \mu }(s_{o})+\int_{s_{o}}^{s}d\overline{s}\left[ \frac{%
\partial g_{A}^{\mu }(r^{\prime })}{\partial r^{\prime \nu }}+A_{(g)\nu
}^{\mu }(r^{\prime },r)\right] u^{\prime \nu }(\overline{s}),
\end{equation}%
where\ manifestly $\int_{s_{o}}^{s}d\overline{s}\frac{\partial g_{A}^{\mu
}(r^{\prime })}{\partial r^{\prime \nu }}u^{\prime \nu }(\overline{s}%
)=g_{A}^{\mu }(r^{\prime }(s))-g_{A}^{\mu }(r^{\prime }(s_{o})).$ Now we
notice that it is always possible to set the initial condition so that $%
r^{\prime \mu }(s_{o})=g_{A}^{\mu }(r^{\prime }(s_{o})).$ This implies the
validity of the first of the Eqs.(\ref{g-NLPT-2}). The proof of the second
one is analogous. \textbf{Q.E.D.}

\bigskip

Notice that, in difference with Eqs.(\ref{NLPT}), the transformations (\ref%
{g-NLPT-2}) (or equivalent (\ref{g-NLPT})) now establish a diffeomorphism
between two \emph{different, connected and time-oriented }space-times $(%
\mathbf{Q}^{4},g)$ and\textbf{\ }$(\mathbf{Q}^{\prime 4},g^{\prime })$ under
the following assumptions:

\begin{itemize}
\item A1) $(\mathbf{Q}^{4},g)$ is an\emph{\ arbitrary curved space-time}$;$

\item A2) $(\mathbf{Q}^{\prime 4},g^{\prime })$ is an\emph{\ arbitrary
curved space-time}$;$

\item B1) the space-times $(\mathbf{Q}^{4},g)$ and $(\mathbf{Q}^{\prime
4},g^{\prime })$ are referred to as \emph{arbitrary} GR-frames;

\item B2) the same space-times $(\mathbf{Q}^{4},g)$ and $(\mathbf{Q}^{\prime
4},g^{\prime })$ are referred to as \emph{different }GR-frames.
\end{itemize}

Let us consider possible particular realizations of the general-NLPT given
above.

The\ first on is obtained dropping assumption B2), i.e., requiring that the
GR-frames of the two space-times $(\mathbf{Q}^{4},g)$ and $(\mathbf{Q}%
^{\prime 4},g^{\prime })$ coincide. In fact, if the coordinate systems for $(%
\mathbf{Q}^{4},g)$ and $(\mathbf{Q}^{\prime 4},g^{\prime })$ are the same
ones while still remaining arbitrary, then one obtains that the constraint
equations%
\begin{eqnarray}
g_{A}^{\mu }(r^{\prime }) &=&r^{\prime \mu },  \notag \\
f_{A}^{\mu }(r) &=&r^{\mu },  \label{requirements for special-NLTP}
\end{eqnarray}%
must hold identically.\textbf{\ }In such a case, denoting the
transformations matrices as%
\begin{eqnarray}
A_{(g)\nu }^{\mu } &=&A_{\nu }^{\mu },  \notag \\
B_{(g)\nu }^{\mu } &=&B_{\nu }^{\mu },  \label{SPECIAL NLPT}
\end{eqnarray}%
the transformations (\ref{g-NLPT-2}) recover the same form given by Eqs.(39)
and (40) in Part 1. These can be conveniently written as%
\begin{equation}
\left\{
\begin{array}{c}
P_{g}\text{: }r^{\mu }(s)=r^{\prime \mu }(s)+\Delta r^{\prime \mu }(s)), \\
P_{g}^{-1}\text{: }r^{\prime \mu }(s)=r^{\mu }(s)+\Delta r^{\mu }(s)),%
\end{array}%
\right.  \label{special-NLPT}
\end{equation}%
with $\Delta r^{\prime \mu }(s))$ and $\Delta r^{\mu }(s)$ identifying the
non-local displacements%
\begin{equation}
\begin{array}{c}
\Delta r^{\prime \mu }(s))=\int_{s_{o}}^{s}d\overline{s}A_{\nu }^{\mu
}(s)u^{\prime \nu }(\overline{s}), \\
\Delta r^{\mu }(s))=\int_{s_{o}}^{s}d\overline{s}B_{\nu }^{\mu }(s)u^{\nu }(%
\overline{s}).%
\end{array}
\label{Delta r(s)}
\end{equation}%
Therefore Eqs.(\ref{g-NLPT-2}) in validity of (\ref{requirements for
special-NLTP}) will be referred to again as \emph{special NLPT. }Their
ensemble realizes manifestly a group (see proof in Part 1), which will be%
\emph{\ }denoted as\emph{\ special NLPT-group }$\left\{ P_{S}\right\} $\emph{%
.} From this conclusion it is immediate to infer the relationship between
general and special NLPT. In fact, it is obvious that for an arbitrary
general NLPT the relationship existing between the Jacobians $M_{(g)\nu
}^{\mu }$ and $M_{\nu }^{\mu },$ as well as the corresponding transformation
matrices $A_{(g)\nu }^{\mu }(r^{\prime },r)$ and $A_{\nu }^{\mu }(r^{\prime
},r),$ is simply provided by the matrix equation%
\begin{equation}
M_{(g)\nu }^{\mu }=M_{\alpha }^{\mu }J_{\nu }^{\alpha },
\end{equation}%
with $J_{\nu }^{\alpha }\equiv \frac{\partial g_{A}^{\alpha }(r^{\prime })}{%
\partial r^{\nu }}$ being the Jacobian of a suitable LPT.

Another interesting realization occurs when the space-time $(\mathbf{Q}%
^{\prime 4},g^{\prime })$ is identified with the Minkowski space-time\textbf{%
\ }$(\mathbf{Q}^{\prime 4},g^{\prime })\equiv (\mathbf{M}^{\prime 4},\eta
^{\prime }(r^{\prime }))$ represented in terms of general curvilinear
coordinates $r^{\prime }\equiv \left\{ r^{\prime \mu }\right\} .$ In such a
case its metric tensor is of the form%
\begin{equation}
\eta _{\mu \nu }^{\prime }(r^{\prime })=J_{\mu }^{\alpha }(r^{\prime
})J_{\nu }^{\beta }(r^{\prime })\eta _{\alpha \beta },
\end{equation}%
with $\eta _{\alpha \beta }$ being the corresponding Minkowski metric tensor
in orthogonal Cartesian coordinates. The corresponding tensor transformation
laws (\ref{implication-2}) become now%
\begin{equation}
\left\{
\begin{tabular}{l}
$g_{\mu \nu }(r)=\left( M_{(g)}^{-1}\right) _{\mu }^{\alpha }\left(
M_{(g)}^{-1}\right) _{\nu }^{\beta }\eta _{\alpha \beta }^{\prime
}(r^{\prime }),$ \\
$\eta _{\mu \nu }^{\prime }(r^{\prime })=M_{(g)\mu }^{\alpha }M_{(g)\nu
}^{\beta }g_{\alpha \beta }(r),$%
\end{tabular}%
\right.  \label{were-1}
\end{equation}%
which generalize those pointed out Part 1 (see Eqs.(42)). However,
remarkably, the corresponding coordinate transformation become now - in
difference to the special NLTP introduced in Part 1 - of the general NLPT
type (\ref{g-NLPT-2}).

It is interesting to stress that the same conclusions, i.e., in particular
Eqs.(\ref{g-NLPT}), can actually be recovered following an alternative
route. This is obtained by introducing suitable prescriptions on the
transformations (\ref{pg1}) and (\ref{pg2}).\textbf{\ }Consider in fact the
following possible realization of the said maps:%
\begin{equation}
\left\{
\begin{array}{c}
r^{\mu }=g_{A}^{\mu }(r^{\prime })+\int_{s_{o}}^{s}d\overline{s}g_{B}^{\mu
}\left( r^{\prime }(\overline{s}),u^{\prime }(\overline{s}),\frac{D^{\prime
}u^{\prime }(\overline{s})}{D\overline{s}}\right) , \\
r^{\prime \mu }=f_{A}^{\mu }(r)+\int_{s_{o}}^{s}d\overline{s}f_{B}^{\mu
}\left( r(\overline{s}),u(\overline{s}),\frac{Du(\overline{s})}{D\overline{s}%
}\right) ,%
\end{array}%
\right.  \label{1B1}
\end{equation}%
where the functions $g_{A}^{\mu }(r^{\prime })$, $g_{B}^{\mu }\left(
r^{\prime }(\overline{s}),u^{\prime }(\overline{s}),\frac{D^{\prime
}u^{\prime }(\overline{s})}{D\overline{s}}\right) $ and $f_{A}^{\mu }(r)$, $%
f_{B}^{\mu }\left( r(\overline{s}),u(\overline{s}),\frac{Du(\overline{s})}{D%
\overline{s}}\right) $ are suitably-defined real and smooth $4-$vector
functions. Notice that by construction Eqs.(\ref{1B1}) are understood as
being evaluated along the corresponding world-lines $r^{\mu }(s)$\ and $%
r^{\prime \mu }(s)$, and therefore they realize a Lagrangian representation
of the NLPT. In particular, let us assume that the $4-$accelerations enter
at most linearly, namely%
\begin{eqnarray}
g_{B}^{\mu }\left( r^{\prime }(\overline{s}),u^{\prime }(\overline{s}),\frac{%
D^{\prime }u^{\prime }(\overline{s})}{D\overline{s}}\right) &\equiv
&G_{k}^{\mu }\frac{D^{\prime }u^{\prime k}(\overline{s})}{D\overline{s}},
\label{ASSUMPT-1} \\
f_{B}^{\mu }\left( r(\overline{s}),u(\overline{s}),\frac{Du(\overline{s})}{D%
\overline{s}}\right) &\equiv &F_{k}^{\mu }\frac{Du^{k}(\overline{s})}{D%
\overline{s}}.  \label{ASSUMPT-2}
\end{eqnarray}%
being $G_{k}^{\mu }$ and $F_{k}^{\mu }$ real functions of the form $%
G_{k}^{\mu }(r^{\prime },\left[ r^{\prime },u^{\prime }\right] )$ and $%
F_{k}^{\mu }(r,\left[ r,u\right] )$ respectively. Next, one notices that
thanks to the validity of the kinematic constraints (\ref{mass-shell
conditions}), the $4-$accelerations $\frac{D^{\prime }u^{\prime k}(\overline{%
s})}{D\overline{s}}$\ and $\frac{Du^{k}(\overline{s})}{D\overline{s}}$ must
necessarily satisfy constraint equations of the type%
\begin{eqnarray}
D^{\prime }u^{\prime \mu } &=&u^{\prime \nu }H_{\nu }^{\prime \mu }ds\equiv
H_{\nu }^{\prime \mu }dr^{\prime \nu },  \label{ASSUMPT-3} \\
Du^{\mu } &=&u^{\nu }H_{\nu }^{\mu }ds\equiv H_{\nu }^{\mu }dr^{\nu },
\label{ASSUMPT-4}
\end{eqnarray}%
with $H_{\nu }^{\prime \mu }$\ and\ $H_{\nu }^{\mu }$\ denoting suitable
antisymmetric tensors, yet to be determined. As a consequence, the
functional form of $g_{B}^{\mu }$ and $f_{B}^{\prime \mu }$ becomes of the
type%
\begin{eqnarray}
g_{B}^{\mu } &=&A_{(g)\nu }^{\mu }\frac{dr^{\prime \nu }}{ds},
\label{ASSUMPT-7} \\
f_{B}^{\prime \mu } &=&B_{(g)\nu }^{\mu }\frac{dr^{\nu }}{ds},
\label{ASSUMPT-8}
\end{eqnarray}%
where the real matrices $A_{\nu }^{\mu }$\ and $B_{\nu }^{\mu }$\ are
defined as%
\begin{eqnarray}
A_{(g)\nu }^{\mu } &=&G_{k}^{\mu }H_{\nu }^{\prime k},  \label{ASSUMPT-5} \\
B_{(g)\nu }^{\mu } &=&F_{k}^{\mu }H_{\nu }^{k}.  \label{ASSUMPT-6}
\end{eqnarray}%
We remark that, despite the matrices $H_{\nu }^{\prime \mu }$\ and\ $H_{\nu
}^{\mu }$\ being anti-symmetric in the upper and lower indices, $A_{(g)\nu
}^{\mu }$\ and $B_{(g)\nu }^{\mu }$\ remain in principle arbitrary, i.e.,
without definite symmetry (or antisymmetry) index properties. In addition,
both matrices $A_{(g)\nu }^{\mu }$\ and $B_{(g)\nu }^{\mu }$ may still
retain both local and non-local functional dependences. Therefore, Eqs.(\ref%
{1B1}) manifestly recover the form (\ref{g-NLPT-2}), i.e., once Eqs.(\ref%
{g-PPG-1}) are invoked in Eq.(\ref{g-NLPT}).

\section{3 - Gedanken experiment interpretation and physical implications of
NLPT}

In this section we analyze certain physical/mathematical implications of the
general NLPT determined by (\ref{g-NLPT-2}) (see THM.1) and the related NLPT
phase-space transformations (\ref{T-PHASE-2}).

The first one concerns the physical interpretation of the NLPT-phase-space
transformation (\ref{T-PHASE-2}) which can be achieved based on the
realization of a GDE. As pointed out in the introduction a possible GDE of
this type is the one which permits to identify the classical dynamical
system (CDS) which is generated by the same phase--space transformation.

The existence of such a CDS is actually immediate. The conclusion follows in
a straightforward way, being in fact analogous to the one displayed in the
Introduction and realized in the context of SR by means of an\textbf{\ }$s-$%
dependent Lorentz boost. For this purpose, let us notice that the NLPT-phase
transformation (\ref{T-PHASE-2}) does indeed generate a CDS.\textbf{\ }In
fact, consider the states $\left\{ r^{\mu }(s),u^{\mu }(s)\right\} $\textbf{%
\ }and $\left\{ r^{\prime \mu }(s),u^{\prime \mu }(s)\right\} $\ involved in
the same transformation (\ref{T-PHASE-2}).

The two maps (\ref{CDS-0}) and (\ref{CDS-1}) are immediately determined
(they are again not independent),\ both being prescribed for all $s_{o},s\in
I.$ This realizes the desired GDE. More precisely: a) the first one, i.e., (%
\ref{CDS-0}) is obtained by considering the state\textbf{\ }$\left\{
r^{\prime \mu }(s),u^{\prime \mu }(s)\right\} $ as a prescribed function of $%
s$ in a suitable interval $I$, so that at all $s$ in the same interval, $%
\left\{ r^{\mu }(s),u^{\mu }(s)\right\} $ is uniquely determined by the same
NLTP transformation; b) the second one represented by Eq.(\ref{CDS-1}) is
obtained instead by considering the state\textbf{\ }$\left\{ r^{\mu
}(s),u^{\mu }(s)\right\} $ as a prescribed function of $s,$ while $\left\{
r^{\prime \mu }(s),u^{\prime \mu }(s)\right\} $ is then determined by the
corresponding NLTP transformation. The two cases a) and b) identify
respectively to the \emph{active }and \emph{passive viewpoints} \emph{for
the same GDE.}

Let us now analyze the conceptual implications of the GDE. For definiteness,
let us assume that the two space-times, namely the \textquotedblleft
current\textquotedblright\ $(\mathbf{Q}^{4},g)$ and the \textquotedblleft
transformed\textquotedblright \textbf{\ }$(\mathbf{Q}^{\prime 4},g^{\prime
}) $ one, are suitably prescribed, together with an arbitrary NLPT
phase-space transformation (\ref{T-PHASE-2}). The active viewpoint of the
same GDE is realized by first assuming that the transformed phase-state
(i.e., the transformed extended GR-frame) $\left\{ r^{\prime \mu
}(s),u^{\prime \mu }(s)\right\} $ is prescribed. This means that $\left\{
r^{\prime \mu }(s),u^{\prime \mu }(s)\right\} $ remains in principle an
arbitrary, but suitably pre-determined, function of $s$. Thus, for example, $%
u^{\prime \mu }(s)$ can always be assumed to be constant for all $s$ in a
prescribed interval $I.$ Then, the GDE permits one to uniquely ideally
\textquotedblleft measure\textquotedblright\ the time-evolution of the state
$\left\{ r^{\mu }(s),u^{\mu }(s)\right\} $ of the the current space-time $(%
\mathbf{Q}^{4},g)$. In the passive viewpoint, instead, the current state
(i.e., the current extended GR-frame) $\left\{ r^{\mu }(s),u^{\mu
}(s)\right\} $ is regarded as prescribed. In this case the GDE permits one
to \textquotedblleft measure\textquotedblleft\ the behavior of the
transformed state $\left\{ r^{\prime \mu }(s),u^{\prime \mu }(s)\right\} $
for the same prescribed NLPT phase-space transformation (\ref{T-PHASE-2}).

Let us now analyze some interesting physical aspects of the theory of NLPT
presented here.

The first one concerns the physical domain of existence of NLTPs. As pointed
out before, just as in the case of LPT, NLTPs must be defined in the
accessible sub-domains of $(\mathbf{Q}^{4},g)$\ and $(\mathbf{Q}^{\prime
4},g^{\prime })$, namely the connected subsets which in each space-time can
be covered by time-like (or if appropriate space-like) world-lines or their
limit functions to be suitably defined. In fact, for example, in the case of
light cones, NLTPs can be defined for time-like world-lines $r^{\mu }(s)$
which are endowed with a $4-$velocity having arbitrarily-large spatial
and/or time components, and therefore arbitrarily close to the same light
trajectories. In addition, we stress that the structure of the two
space-times themselves remains \textquotedblleft a priori\textquotedblright\
arbitrary. Thus, for example, each of them may be characterized by different
ensembles of event horizons, while NLTPs remain defined in the subsets
internal or external to the same event horizons such that the mapped subsets
have the same signature.

A further aspect to be mentioned concerns the tensor transformation laws
with respect to the general NLPT-group $\left\{ P_{g}\right\} .$ Indeed Eqs.(%
\ref{implication-1})-(\ref{implication-2}) are the prototypes\ of tensor
transformations laws which can be extended to virtually arbitrary
higher--rank tensors. Thus, as an illustration, let us consider the case of
a $4-$scalar field $\Phi (r)$,\ i.e., a function which remains invariant
under the action of an arbitrary transformation of the group $\left\{
P_{g}\right\} $,\ for example identified with the special NLPT%
\begin{equation}
r^{\mu }\equiv r^{\mu }(r^{\prime \mu }(s),s)=r^{\prime \mu }(s)+\Delta
r^{\prime \mu }(s),  \label{NL}
\end{equation}%
with\ $\Delta r^{\prime \mu }(s)$\ being defined by Eq.(\ref{Delta r(s)}).%
\textbf{\ }Then, denoting as $\Phi ^{\prime }(r^{\prime })$\ (respectively $%
\Phi (r))$\ the realization of the same scalar field in the GR-reference
frame $r^{\prime \mu }$\ (respectively $r^{\mu }$), it follows that the
Eulerian equation%
\begin{equation}
\Phi ^{\prime }(r^{\prime })=\Phi (r)  \label{frist equation}
\end{equation}%
must hold identically.\textbf{\ }On the other hand, on the rhs of the same
equation $r\equiv \left\{ r^{\mu }\right\} $\ is to be considered a function
of $r^{\prime }\equiv \left\{ r^{\prime \mu }\right\} $\ when represented
via the the special NLPT given above.\textbf{\ }It follows that $\Phi
(r(s))\equiv \Phi (r^{\prime \mu }(s)+\Delta r^{\prime \mu }(s))$\ when cast
in Lagrangian form, i.e., it is parametrized in terms of the world-line $%
r^{\mu }(s)$\ or $r^{\prime \mu }(s)$\ respectively and the corresponding
proper time $s$.\textbf{\ }As a result, Eq.(\ref{frist equation}) yields
also the relationship expressed in Lagrangian form, i.e., in terms of the
world-lines $r(s)$\ and $r^{\prime }(s)$.\textbf{\ }Since by construction $%
r(s)$\ is a non-local function of $r^{\prime }(s)$\ and the initial and
transformed fields $\Phi (r(s))$\ must still coincide identically, i.e.,%
\begin{equation}
\Phi ^{\prime }(r^{\prime }(s))=\Phi (r(s))\equiv \Phi (r^{\prime \mu
}(s)+\Delta r^{\prime \mu }(s)),  \label{second equation}
\end{equation}%
it follows that $\Phi ^{\prime }(r^{\prime }(s))$\ becomes necessarily a
non-local function of $r^{\prime \mu }(s)$. To determine the corresponding
Eulerian fields in terms of Eq.(\ref{frist equation}) it is sufficient to
represent the proper time $s$\ in terms of the instantaneous $4-$position $%
r^{\prime }\equiv \left\{ r^{\prime \mu }\right\} $,\textbf{\ }so that $%
s=s(r^{\prime })$.\textbf{\ }The way how this can be done, once the
world-line $r^{\prime \mu }(s)$\ is considered prescribed, is discussed in
the Appendix. Once the representation $s=s(r^{\prime })$\ is introduced, it
follows that the rhs of Eq.(\ref{second equation}) determines actually a
function of $r^{\prime }\equiv \left\{ r^{\prime \mu }\right\} $\ only,
namely%
\begin{equation}
\Phi (r^{\prime \mu }+\Delta r^{\prime \mu }(s))\equiv \widehat{\Phi }%
(r^{\prime }),
\end{equation}%
so that Eq.(\ref{frist equation}) implies%
\begin{equation}
\Phi ^{\prime }(r^{\prime })\equiv \widehat{\Phi }(r^{\prime })
\end{equation}%
too. In other words, the scalar field $\Phi (r)$\ and hence$\ \Phi ^{\prime
}(r^{\prime })$\ become formally a composite and non-local function of $%
r^{\prime }\equiv \left\{ r^{\prime \mu }\right\} $.

Finally, a number of comments and suggestions related to the form of the
general NLPT, realized in particular in THM.1 and in the subsequent
discussion, should be mentioned. These include:

1) The two matrices $A_{(g)\nu }^{\mu }\left( r^{\prime },r\right) $ and $%
B_{(g)\nu }^{\mu }\left( r,r^{\prime }\right) $ identify the
acceleration-dependent contributions in the Jacobian matrices.

2)\ It must be stressed that the two involved metric tensors $g_{\mu \nu }$
and $g_{\mu \nu }^{\prime }$ remain arbitrary. For example, one can always
require that both metric tensors are particular solutions of the Einstein
equation. In this case Eqs. (\ref{implication-2}) can be interpreted as
equations for the still unknown Jacobian matrix, to be determined
accordingly. This includes as a particular case the one in which for example
the transformed metric tensor $g_{\alpha \beta }^{\prime }\left( r^{\prime
}\right) $ coincides with the Minkowski metric tensor. If $g_{\mu \nu
}\left( r\right) $ and $g_{\mu \nu }^{\prime }\left( r^{\prime }\right) $
are realizations holding for the two different space-times $(\mathbf{Q}%
^{4},g)$\ and $(\mathbf{Q}^{\prime 4},g^{\prime })$ when they are referred
respectively to the coordinate systems $r^{\mu }$ and $r^{\prime \mu }$, the
tensor transformation laws (\ref{implication-2}) must hold.\ If the vector
functions $g_{A}^{\mu }(r^{\prime })$ and $f_{A}^{\mu }(r)$ are considered
prescribed, then the first of these equations becomes%
\begin{equation}
g_{\mu \nu }\left( r\right) =\left[ \frac{\partial g_{A}^{\alpha }(r^{\prime
})}{\partial r^{\prime \mu }}+A_{(g)\mu }^{\alpha }\left( r^{\prime
},r\right) \right] \left[ \frac{\partial g_{A}^{\beta }(r^{\prime })}{%
\partial r^{\prime \nu }}+A_{(g)\nu }^{\beta }\left( r^{\prime },r\right) %
\right] g_{\alpha \beta }^{\prime }\left( r^{\prime }\right) ,
\label{PROBLEM-1}
\end{equation}%
which, for special NLPT (see for example Eqs.(\ref{special-NLPT})), reduces
simply to%
\begin{equation}
g_{\mu \nu }\left( r\right) =\left[ \delta _{\mu }^{\alpha }+A_{(g)\mu
}^{\alpha }\left( r^{\prime },r\right) \right] \left[ \delta _{\nu }^{\beta
}+A_{(g)\nu }^{\beta }\left( r^{\prime },r\right) \right] g_{\alpha \beta
}^{\prime }\left( r^{\prime }\right) .  \label{PROBLEM-2}
\end{equation}%
Eq.(\ref{PROBLEM-1}), or alternatively (\ref{PROBLEM-2}), yields actually a
set of implicit, i.e., integral, equations for the components of the same
matrix. The explicit construction of the solution for $A_{(g)\nu }^{\beta }$
actually requires representing it in Eulerian form. This involves as before
(see related discussion in the previous section) representing the
proper-time $s$ in terms of the instantaneous $4-$position $r^{\prime
}\equiv \left\{ r^{\prime \mu }\right\} $, so that $s=s(r^{\prime })$. We
refer again for this purpose to the discussion reported in the Appendix.

3) An alternative interpretation is the one in which one of the two metric
tensors, say $g_{\alpha \beta }^{\prime }\left( r^{\prime }\right) $, is
prescribed together\ with the Jacobian\ $M_{\mu }^{\alpha }\left( r^{\prime
}\right) $ so that Eqs.(\ref{implication-2}) provides an explicit
representation for the transformed metric tensor $g_{\mu \nu }\left(
r\right) $. In this case an interesting remaining issue concerns its
possible identification as as an admissible particular solution of the
Einstein equation corresponding to prescribed physical sources.

4)\ The problem of the construction of the NLPT - or, better, the
corresponding special NLPT to which in principle it should always be
possible to refer - amounts therefore to look for the still unknown matrix $%
A_{(g)\nu }^{\beta }\left( r^{\prime },r\right) $.

\section{4 - Application \#1: Diagonal metric tensors}

The first application to be considered concerns the construction of a NLPT
mapping two connected and time-oriented space-times $(\mathbf{Q}^{4},g)$\
and $(\mathbf{Q}^{\prime 4},g^{\prime })$ both having diagonal form with
respect to suitable sets of coordinates. More precisely we shall require
that:

\begin{itemize}
\item When $(\mathbf{Q}^{4},g)$\ and $(\mathbf{Q}^{\prime 4},g^{\prime })$
are referred to the same coordinate systems, both are realized by diagonal
metric tensors%
\begin{equation}
\left\{
\begin{array}{c}
g_{\mu \nu }(r)\equiv diag\left(
S_{0}(r),-S_{1}(r),-S_{2}(r),-S_{3}(r)\right) \\
g_{\mu \nu }^{\prime }(r^{\prime })\equiv diag\left( S_{0}^{\prime
}(r^{\prime }),-S_{1}^{\prime }(r^{\prime }),-S_{2}^{\prime }(r^{\prime
}),-S_{3}^{\prime }(r^{\prime })\right)%
\end{array}%
\right.  \label{diagonal metric tensors}
\end{equation}%
\ respectively.\ The accessible subsets are as follows: a) for $(\mathbf{Q}%
^{\prime 4},g^{\prime })$ is that in which for all $\mu =0,3,$ $S_{\mu
}^{\prime }(r^{\prime })>0;$ b) for $(\mathbf{Q}^{4},g)$ is either the set
in which for all $\mu =0,3,$ $S_{\mu }(r)>0$ or the other one in which $%
S_{0}(r)<0,S_{1}(r)<0,S_{2}(r)>0$ and $S_{3}(r)>0.$

\item $(\mathbf{Q}^{4},g)$\ and $(\mathbf{Q}^{\prime 4},g^{\prime })$ are
intrinsically \emph{different}, i.e., that the corresponding Riemann
curvature tensors $R_{\mu \nu }(r)$\ and $R_{\mu \nu }^{\prime }(r^{\prime
}) $\ cannot be globally mapped in each other by means of any LPT. This
means that a mapping between the accessible subsets of the said space-times
can only possibly be established by means of a suitable NLPT.

\item Two occurrences are considered: a) the \emph{same-signature} case in
which both $(\mathbf{Q}^{\prime 4},g^{\prime })$ and $(\mathbf{Q}^{4},g)$
have the same Lorentzian signature $(+,-,-,-);$ b) the \emph{%
opposite-signature} case in which $(\mathbf{Q}^{\prime 4},g^{\prime })$ and $%
(\mathbf{Q}^{4},g)$ have signatures $(+,-,-,-)$ and $(-,+,+,+)$ respectively$%
.$
\end{itemize}

In validity of Eqs.(\ref{diagonal metric tensors}) the tensor transformation
equation (\ref{implication-2}) take obviously the general form:%
\begin{equation}
\left\{
\begin{tabular}{l}
$S_{\mu }(r)=\left( M_{(g)}^{-1}\right) _{\mu }^{\alpha }(r,r^{\prime
})\left( M_{(g)}^{-1}\right) _{(\mu )}^{\alpha }(r,r^{\prime })S_{\alpha
}^{\prime }(r^{\prime }),$ \\
$S_{\mu }^{\prime }(r^{\prime })=M_{(g)\mu }^{\alpha }(r^{\prime
},r)M_{(g)(\mu )}^{\alpha }(r^{\prime },r)S_{\alpha }(r),$%
\end{tabular}%
\right.  \label{diagonal-case-SOLUTION}
\end{equation}%
where manifestly $M_{(g)\mu }^{\alpha }(r^{\prime },r)\equiv M_{\mu
}^{\alpha }(r^{\prime },r)$ and $\left( M_{(g)}^{-1}\right) _{\mu }^{\alpha
}(r,r^{\prime })=\left( M^{-1}\right) _{\mu }^{\alpha }(r,r^{\prime })$ as
corresponds to the case of a special NLTP. For such a type of space-times in
the following we intend to display a number of explicit particular solutions
of Eqs.(\ref{diagonal-case-SOLUTION}) for the Jacobian $M_{\mu }^{\alpha }$
and its inverse $\left( M^{-1}\right) _{\mu }^{\alpha },$ and to construct
also the corresponding NLPT-phase-space transformations.

\subsection{Same-signature diagonal NLTP}

In the case in which $(\mathbf{Q}^{4},g)$\ and $(\mathbf{Q}^{\prime
4},g^{\prime })$ have the same signatures, it is immediate to show that a
particular solution of Eqs.(\ref{diagonal-case-SOLUTION}) in the accessible
subsets of $(\mathbf{Q}^{4},g)$\ and $(\mathbf{Q}^{\prime 4},g^{\prime })$
is provided by a diagonal Jacobian matrix, i.e., of the form%
\begin{equation}
M_{\mu }^{\alpha }(r^{\prime },r)=M_{\mu }^{\mu }(r^{\prime },r)\delta _{\mu
}^{\alpha }\equiv \left[ \delta _{\mu }^{\alpha }+A_{\mu }^{\mu }(r^{\prime
},r)\right] \delta _{\mu }^{\alpha }.  \label{DIAGONAL MATRIX}
\end{equation}%
Indeed from Eqs.(\ref{diagonal-case-SOLUTION}) one finds
\begin{equation}
M_{(\mu )}^{\mu }(r^{\prime },r)=\frac{1}{\left( M^{-1}\right) _{(\mu
)}^{\mu }}=\sqrt{\frac{S_{\mu }^{\prime }(r^{\prime })}{S_{(\mu )}(r)}},
\end{equation}%
where $\frac{S_{\mu }^{\prime }(r^{\prime })}{S_{(\mu )}(r)}>0$ in the
accessible subsets. In terms of Eqs.(\ref{NLPT}), or equivalent (\ref{g-NLPT}%
), one then determines the corresponding special NLPT, namely%
\begin{equation}
\left\{
\begin{array}{c}
\text{ }r^{\mu }(s)=r^{\prime \mu }(s_{o})+\int_{s_{o}}^{s}d\overline{s}%
\sqrt{\frac{S_{\mu }^{\prime }(r^{\prime })}{S_{(\mu )}(r)}}u^{\prime \mu }(%
\overline{s}), \\
\text{ }r^{\prime \mu }(s)=r^{\mu }(s_{o})+\int_{s_{o}}^{s}d\overline{s}%
\sqrt{\frac{S_{(\mu )}(r^{\prime })}{S_{\mu }^{\prime }(r^{\prime })}}u^{\mu
}(\overline{s}),%
\end{array}%
\right.
\end{equation}%
as well as the corresponding $4-$ velocity transformation.

Let us now consider a possible physical realizations for the space-times $(%
\mathbf{Q}^{4},g)$ and $(\mathbf{Q}^{\prime 4},g^{\prime })$ and the
corresponding metric tensors $g_{\mu \nu }(r)$\ and $g_{\mu \nu }^{\prime
}(r^{\prime })$ respectively. Examples are provided by the Schwarzschild or
alternatively the Reissner-Nordstr\"{o}m space-times, both being
characterized by a single event horizon. In terms of the spherical
coordinates $\left( r,\vartheta ,\varphi \right) $ an analogous (\textit{%
Schwarzschild-analog}) representation holds of the form $g_{\mu \nu
}(r)\equiv $diag$\left( \left( S_{0}(r),-S_{1}(r),-S_{2}(r),-S_{3}(r)\right)
\right) $ with%
\begin{equation}
\left\{
\begin{array}{c}
S_{0}(r)=f(r) \\
S_{1}(r)=\frac{1}{f(r)} \\
S_{2}(r)=r^{2} \\
S_{3}(r)=r^{2}\sin ^{2}\vartheta%
\end{array}%
\right. ,  \label{SCHW-1}
\end{equation}%
and where in the two cases $f(r)$ is identified respectively with%
\begin{eqnarray}
f(r) &=&\left( 1-\frac{r_{s}}{r}\right) ,  \label{schwartz} \\
f(r) &=&\left( 1-\frac{r_{s}}{r}+\frac{r_{Q}^{2}}{r^{2}}\right) .  \label{RN}
\end{eqnarray}%
Here, $r_{s}=2GM/c^{2}$\ is the Schwarzschild radius and $r_{Q}=\sqrt{\frac{%
Q^{2}G}{4\pi \varepsilon _{0}c^{4}}}$\ a characteristic length scale, with $%
Q $\ being the electric charge and $1/4\pi \varepsilon _{0}$\ the Coulomb
coupling constant. Introducing the curvilinear coordinates $\left(
r^{0},r^{1}\equiv r,r^{2}\equiv r\vartheta ,r^{3}\equiv \varphi r\sin
\vartheta \right) $,\ here referred to as pseudo-spherical coordinates,\ one
obtains $r^{2}d\Omega ^{2}=\left( dr^{2}\right) ^{2}+\left( dr^{3}\right)
^{2}$. It follows that in Eqs.(\ref{schwartz}) and (\ref{RN}), $S_{2}(r)$\
and $S_{3}(r)$\ are replaced with%
\begin{eqnarray}
S_{2}(r) &=&1, \\
S_{3}(r) &=&1.
\end{eqnarray}%
In both cases, the transformed space-time $(\mathbf{Q}^{\prime 4},g^{\prime
})$\ is assumed again Schwarzschild-analog,\ namely\ of the type (\ref%
{SCHW-1}). Expressed in the pseudo-spherical coordinates this is prescribed
to be%
\begin{equation}
\left\{
\begin{array}{c}
S_{0}^{\prime }(r^{\prime })=f^{\prime }(r^{\prime }) \\
S_{1}^{\prime }(r^{\prime })=\frac{1}{f^{\prime }(r^{\prime })} \\
S_{2}^{\prime }(r^{\prime })=1 \\
S_{3}^{\prime }(r^{\prime })=1%
\end{array}%
\right. .  \label{scw-2}
\end{equation}%
Here $f^{\prime }(r^{\prime })$\ is assumed to be an analytic function
having $n>1$\ positive simple roots $r_{1}^{\prime }<r_{2}^{\prime
}<...<r_{n}^{\prime }$\ in the positive real axis $\left[ 0,+\infty \right] $
and such that $f^{\prime }(r^{\prime })>0$\ for $r^{\prime }>r_{n}^{\prime }$%
. In particular, we shall require that the Schwarzschild radius occurs in
the interval%
\begin{equation}
r_{1}^{\prime }<r_{s}<r_{n}^{\prime }.
\end{equation}%
The admissible sub-domains of $(\mathbf{Q}^{4},g)$\textbf{\ }and\textbf{\ }$(%
\mathbf{Q}^{\prime 4},g^{\prime })$, where NLPTs can possibly be established
between the two space-times, are therefore defined respectively by the
inequalities $r>r_{s}$ and $r^{\prime }>r_{n}^{\prime }$. In these subsets
the transformation matrix $A_{\mu }^{\nu }(r^{\prime },r)$ becomes:%
\begin{eqnarray}
A_{0}^{0}(r^{\prime },r) &=&\sqrt{\frac{f^{\prime }(r^{\prime })}{f(r)}}-1,
\label{acc-1} \\
A_{1}^{1}(r^{\prime },r) &=&\sqrt{\frac{f(r)}{f^{\prime }(r^{\prime })}}-1,
\label{acc-2} \\
A_{2}^{2}(r^{\prime },r) &=&\sqrt{\frac{1}{1}}-1=0,  \label{acc-3} \\
A_{3}^{3}(r^{\prime },r) &=&\sqrt{\frac{1}{1}}-1=0,  \label{acc-4}
\end{eqnarray}%
where in the first terms on the rhs of the previous equations the positive
values of the square roots have been taken. Therefore, the NLPT\textbf{\ }%
corresponding to Eqs.(\ref{acc-1})-(\ref{acc-4}) is the identity
transformation as far as the coordinates $r^{2}$ and $r^{3}$ are concerned.
The non-trivial contributions giving rise to non-local terms in Eqs. (\ref%
{g-NLPT}) are produced therefore only by the time and radial components of
the\textbf{\ }$4-$velocity, i.e., $u^{\prime 0}$ and $u^{\prime 1}$ only.
The following physical interpretation is proposed:

\begin{itemize}
\item The special NLPT corresponding to Eqs.(\ref{acc-1})-(\ref{acc-4}) is
only defined in the accessible subset of the space-timed namely when $%
r^{\prime }>r_{n}^{\prime }$ and $r>r_{s\text{.}}$respectively occur.

\item The effect of the special NLPT produced by Eqs.(\ref{acc-1})-(\ref%
{acc-4}) is that of mapping the accessible subsets of Schwarzschild or
Reissner-Nordstr\"{o}m space-time in the corresponding accessible subset of
a Schwarzschild-analog space-time. The basic feature of the transformed
space-time is that of exhibiting $n>1$ event-horizons instead of a single
one as in the initial space-time.

\item The physical origin for the generation of such an effect is the
special NLPT introduced here, which in turn arises when non-local effects
are included in Eq.(\ref{NLPT}) which are carried only by the time and
radial components of the\textbf{\ }$4-$velocity. In particular, assuming
that the NLPT is of the form determined according to the requirements\textbf{%
\ }(\ref{ASSUMPT-1}) it follows that Eqs.(\ref{acc-1})-(\ref{acc-4})
correspond to the case in which only a tangential $4-$acceleration $%
a^{\prime \mu }=\frac{D^{\prime }u^{\prime \mu }}{Ds}$\ can occur, namely in
which its only non-vanishing components correspond to $\mu =2,3$.
\end{itemize}

A final remark must be made concerning the limit $\lim_{r^{\prime
}\rightarrow r_{\ast }^{\prime (+)}}$ in Eq.(\ref{acc-2}) and respectively $%
\lim_{r\rightarrow r_{\ast }^{(+)}}$ in Eqs.(\ref{acc-1}), where $r_{\ast
}^{\prime }$ and $r_{\ast }$ are the largest roots of the equations $%
f^{\prime }(r^{\prime })=0$ and $f(r)=0$. In terms of the pseudo-spherical
coordinates the previous limits do not exist and therefore the limit NLPT is
not defined on the event horizons. Nevertheless, these divergences can be
cured by preliminarily recurring to a suitable coordinate system, which in
the case of the Schwarzschild metric can be identified with the
Kruskal--Szekeres coordinates \cite{Wheeler}.

\subsection{Opposite-signature NLTP}

Let us now consider the case in which $(\mathbf{Q}^{4},g)$\ and $(\mathbf{Q}%
^{\prime 4},g^{\prime })$ have opposite signatures, namely respectively $%
(-,+,+,+)$ and $(+,-.-,-)$ while the metric tensors are still diagonal when
expressed with respect to the same coordinate systems, i.e., are of the form
(\ref{diagonal metric tensors}). It follows that in the accessible subset of
$(\mathbf{Q}^{\prime 4},g^{\prime })$ it occurs respectively that%
\begin{eqnarray}
S_{0}(r) &<&0,  \notag \\
S_{1}(r) &<&0.
\end{eqnarray}%
In this case it is immediate to show that in the accessible subsets of $(%
\mathbf{Q}^{4},g)$\ and $(\mathbf{Q}^{\prime 4},g^{\prime })$ a particular
solution of Eqs.(\ref{diagonal-case-SOLUTION}) is provided by a Jacobian
matrix of the form
\begin{eqnarray}
M_{1}^{0}(r^{\prime },r) &=&\frac{1}{\left( M^{-1}\right) _{0}^{1}}=\sqrt{-%
\frac{S_{1}^{\prime }(r^{\prime })}{S_{0}(r)}}, \\
M_{0}^{1}(r^{\prime },r) &=&\frac{1}{\left( M^{-1}\right) _{1}^{0}}=\sqrt{-%
\frac{S_{0}^{\prime }(r^{\prime })}{S_{1}(r)}}, \\
M_{2}^{2}(r^{\prime },r) &=&\frac{1}{\left( M^{-1}\right) _{3}^{3}}=\sqrt{%
\frac{S_{2}^{\prime }(r^{\prime })}{S_{2}(r)}}, \\
M_{3}^{3}(r^{\prime },r) &=&\frac{1}{\left( M^{-1}\right) _{3}^{3}}=\sqrt{%
\frac{S_{3}^{\prime }(r^{\prime })}{S_{3}(r)}},
\end{eqnarray}%
where $-\frac{S_{1}(r)}{S_{0}^{\prime }(r^{\prime })}>0$ and $-\frac{S_{0}(r)%
}{S_{1}^{\prime }(r^{\prime })}$ in the accessible subsets. The
corresponding special NLPT follows immediately from Eqs.(\ref{NLPT}), or
equivalent (\ref{g-NLPT}). Once again a possible application is provided by
Schwarzschild-analo\textit{g} space-times. More precisely let us consider
the case in which:

A) the space-time $(\mathbf{Q}^{\prime 4},g^{\prime })$ is assumed again
Schwarzschild-analo\textit{g} of the type (\ref{SCHW-1}), so that in
pseudo-spherical coordinates it is given again by Eqs.(\ref{scw-2}). In
particular in the accessible subset of $(\mathbf{Q}^{\prime 4},g^{\prime })$
we shall require%
\begin{equation}
\left\{
\begin{array}{c}
S_{0}^{\prime }(r^{\prime })=f^{\prime }(r^{\prime })>0 \\
S_{1}^{\prime }(r^{\prime })=\frac{1}{f^{\prime }(r^{\prime })}>0%
\end{array}%
\right. .
\end{equation}

B) the space-time $(\mathbf{Q}^{4},g)$ is the Schwarzschild one, the
accessible subset being such that%
\begin{equation}
\left\{
\begin{array}{c}
S_{0}(r)=1-\frac{r_{s}}{r}<0 \\
S_{1}(r)=\frac{1}{1-\frac{r_{s}}{r}}<0%
\end{array}%
\right. .
\end{equation}%
As a consequence, the Jacobian\ becomes%
\begin{eqnarray}
M_{1}^{0}(r^{\prime },r) &=&\frac{1}{\left( M^{-1}\right) _{0}^{1}}=\sqrt{-%
\frac{1}{\left( 1-\frac{r_{s}}{r}\right) f^{\prime }(r^{\prime })}}, \\
M_{0}^{1}(r^{\prime },r) &=&\frac{1}{\left( M^{-1}\right) _{1}^{0}}=\sqrt{%
-\left( 1-\frac{r_{s}}{r}\right) f^{\prime }(r^{\prime })}, \\
M_{2}^{2}(r^{\prime },r) &=&\frac{1}{\left( M^{-1}\right) _{3}^{3}}=1, \\
M_{3}^{3}(r^{\prime },r) &=&\frac{1}{\left( M^{-1}\right) _{3}^{3}}=1.
\end{eqnarray}

Therefore, in this case the resulting special NLPT maps the interior domain
of the Schwarzschild space-time, namely its \emph{Black Hole domain}, onto
the exterior domain of a Schwarzschild-analog space-time. As a final
comment, it must be stressed that the starting equations adopted in this
Section, namely Eqs.(\ref{DIAGONAL MATRIX}), can be in principle easily
reformulated when arbitrary different coordinate systems are adopted for
representing the two space-times $(\mathbf{Q}^{4},g)$\textbf{\ }and $(%
\mathbf{Q}^{\prime 4},g^{\prime })$. Although details are here omitted for
brevity, it is worth mentioning that this extension can easily be
accomplished adopting the general NLPT- theory developed here.

\section{5 - Application \#2: Diagonalization of metric tensors}

As a second example, the problem of diagonalization of a non-diagonal metric
tensor is posed in the framework of NLPT-theory. More precisely, this
concerns the construction of a NLPT mapping two connected and time-oriented
space-times $(\mathbf{Q}^{4},g)$\ and $(\mathbf{Q}^{\prime 4},g^{\prime })$.
Here we shall require that when $(\mathbf{Q}^{4},g)$\ and $(\mathbf{Q}%
^{\prime 4},g^{\prime })$ are referred to the same coordinate systems they
are realized by the metric tensors
\begin{equation}
g_{\mu \nu }(r)\equiv diag\left(
S_{0}(r),-S_{1}(r),-S_{2}(r),-S_{3}(r)\right) ,
\end{equation}%
\begin{equation}
g_{\mu \nu }^{\prime }(r^{\prime })=\left\vert
\begin{array}{cccc}
S_{0}^{\prime }(r^{\prime }) &  &  & S_{03}^{\prime }(r^{\prime }) \\
& -S_{1}^{\prime }(r^{\prime }) &  &  \\
&  & -S_{2}^{\prime }(r^{\prime }) &  \\
S_{03}^{\prime }(r^{\prime }) &  &  & -S_{3}^{\prime }(r^{\prime })%
\end{array}%
\right\vert ,
\end{equation}%
\ respectively.\ The accessible subsets are assumed to be both for $(\mathbf{%
Q}^{\prime 4},g^{\prime })$ and $(\mathbf{Q}^{4},g)$ as follows: all $\mu
=0,3,$ $S_{\mu }^{\prime }(r^{\prime })>0$ and $S_{\mu }(r)>0$ $.$

As before, the realization of the NLPT which maps the two metric tensors is
not unique. A possible choice is provided by a special NLPT of the form%
\begin{eqnarray}
dr^{0} &=&\left( 1+A_{(g)0}^{0}\right) dr^{\prime 0}+A_{(g)3}^{0}dr^{\prime
3},  \label{po-1} \\
dr^{i} &=&\left( 1+A_{(g)\left( i\right) }^{i}\right) dr^{\prime \left(
i\right) },  \label{po-2}
\end{eqnarray}%
for $i=1,2,3$, namely such that%
\begin{eqnarray}
r^{0}(s) &=&r^{\prime 0}(s)+\int_{s_{o}}^{s}d\overline{s}\left[
A_{(g)0}^{0}(r^{\prime }(\overline{s}),r(\overline{s}))u^{\prime 0}(%
\overline{s})+A_{(g)3}^{0}(r^{\prime }(\overline{s}),r(\overline{s}%
))u^{\prime 3}(\overline{s})\right] ,  \label{nd-int-1} \\
r^{i}(s) &=&r^{\prime i}(s)+\int_{s_{o}}^{s}d\overline{s}A_{(g)\left(
i\right) }^{i}(r^{\prime }(\overline{s}),r(\overline{s}))u^{\prime \left(
i\right) }(\overline{s}),  \label{nd-int-2}
\end{eqnarray}%
where again the indices in brackets are not subject to the summation rule.
The transformation bringing $r^{\prime \mu }$ in $r^{\mu }$ will be referred
to as \emph{diagonalizing NLPT}. The transformation equations for the matrix
elements $A_{(g)0}^{0}$, $A_{(g)3}^{0}$ and $A_{(g)\left( i\right) }^{i}$,
for $i=1,2,3$, are therefore%
\begin{eqnarray}
S_{j}^{\prime }\left( r^{\prime }\right) &=&\left( 1+A_{(g)\left( j\right)
}^{\left( j\right) }(r^{\prime },r)\right) ^{2}S_{j}\left( r\right) ,
\label{nd-1} \\
S_{3}^{\prime }\left( r^{\prime }\right) &=&\left( 1+A_{(g)3}^{3}(r^{\prime
},r)\right) ^{2}S_{3}\left( r\right) -\left( A_{(g)3}^{0}(r^{\prime
},r)\right) ^{2}S_{0}\left( r\right) ,  \label{nd-2} \\
S_{03}^{\prime }\left( r^{\prime }\right) &=&A_{(g)3}^{0}(r^{\prime
},r)A_{(g)0}^{0}(r^{\prime },r)S_{0}\left( r\right) ,  \label{nd-3}
\end{eqnarray}%
for $j=0,1,2$. The first set of equations (\ref{nd-1}) has a formal solution
of the type%
\begin{equation}
A_{(g)\left( j\right) }^{j}(r^{\prime },r)=\sqrt{\frac{S_{j}^{\prime }\left(
r^{\prime }\right) }{S_{\left( j\right) }\left( r\right) }}-1.  \label{xxx-1}
\end{equation}%
The third equation (\ref{nd-3}) gives then%
\begin{equation}
A_{(g)3}^{0}(r^{\prime },r)=\frac{S_{03}^{\prime }\left( r^{\prime }\right)
}{S_{0}\left( r\right) }\left[ \sqrt{\frac{S_{0}^{\prime }\left( r^{\prime
}\right) }{S_{0}\left( r\right) }}-1\right] ^{-1}.  \label{xxx-2}
\end{equation}%
Finally, Eq.(\ref{nd-2}) delivers%
\begin{equation}
A_{(g)3}^{3}(r^{\prime },r)=\sqrt{\frac{S_{3}^{\prime }\left( r^{\prime
}\right) +\left( A_{3}^{0}(r^{\prime },r)\right) ^{2}S_{0}\left( r\right) }{%
S_{3}\left( r\right) }}-1.  \label{xxx-3}
\end{equation}%
The signs of the square roots in the previous equations have been chosen in
such a way to recover the correct result for identity transformations.

A number of remarks must be made:

1) Also the present application can be in principle reformulated adopting
arbitrary different coordinate systems for the representation of the
space-times $(\mathbf{Q}^{4},g)$\ and $(\mathbf{Q}^{\prime 4},g^{\prime })$.
This ultimately involves adopting the general NLPT- theory developed here.

2)\ The transformation (\ref{nd-int-1})-(\ref{nd-int-2}) is defined provided
the inequality%
\begin{equation}
\sqrt{\frac{S_{0}^{\prime }\left( r^{\prime }\right) }{S_{0}\left( r\right) }%
}-1\neq 0  \label{nd-ineq}
\end{equation}%
holds. In this case in fact all the matrix elements $A_{\nu }^{\mu }$
determined above are real and smooth functions.

3)\ A solution satisfying the inequality (\ref{nd-ineq}) can always be found
by suitably prescribing $S_{0}\left( r\right) $ once $S_{0}^{\prime }\left(
r^{\prime }\right) $ is considered fixed.

4) As an alternate possibility, in case the condition (\ref{nd-ineq}) is not
satisfied, is to look for another possible realization of the transformation
(\ref{nd-int-1})-(\ref{nd-int-2}). The general solution can be cast in the
form%
\begin{eqnarray}
dr^{0} &=&\left( 1+A_{(g)0}^{0}\right) dr^{\prime 0}+A_{(g)3}^{0}dr^{\prime
3}, \\
dr^{i} &=&\left( 1+A_{(g)\left( i\right) }^{i}\right) dr^{\prime \left(
i\right) }, \\
dr^{3} &=&\left( 1+A_{(g)3}^{3}\right) dr^{\prime 3}+A_{(g)0}^{3}dr^{\prime
0},
\end{eqnarray}%
for $i=1,2$, namely such that%
\begin{eqnarray}
r^{0}(s) &=&r^{\prime 0}(s)+\int_{s_{o}}^{s}d\overline{s}\left[
A_{(g)0}^{0}(r^{\prime }(\overline{s}),r(\overline{s}))u^{\prime 0}(%
\overline{s})+A_{(g)3}^{0}(r^{\prime }(\overline{s}),r(\overline{s}%
))u^{\prime 3}(\overline{s})\right] ,  \label{wer-1} \\
r^{i}(s) &=&r^{\prime i}(s)+\int_{s_{o}}^{s}d\overline{s}A_{(g)\left(
i\right) }^{i}(r^{\prime }(\overline{s}),r(\overline{s}))u^{\prime \left(
i\right) }(\overline{s}), \\
r^{3}(s) &=&r^{\prime 3}(s)+\int_{s_{o}}^{s}d\overline{s}\left[
A_{(g)3}^{3}(r^{\prime }(\overline{s}),r(\overline{s}))u^{\prime 3}(%
\overline{s})+A_{(g)0}^{3}(r^{\prime }(\overline{s}),r(\overline{s}%
))u^{\prime 0}(\overline{s})\right] .  \label{wer-3}
\end{eqnarray}%
The resulting equations can be immediately solved.

5)\ The diagonalization of the Kerr metric tensor expressed in spherical
coordinates, as well as the Kerr-Newman and analogous Kerr-like solutions,
can be carried out in terms of either a transformation of the type (\ref%
{nd-int-1})-(\ref{nd-int-2}) or (\ref{wer-1})-(\ref{wer-3}).

6)\ Regarding the physical interpretation of the differential equations (\ref%
{po-1})-(\ref{po-2}) we notice that the first equation implies that the time
component of the 4-velocity in the initial frame is modified by the combined
effects of time- and 3-components of the 4-velocity in the transformed
frame. In the case of the Kerr metric, in particular, the latter corresponds
to an azimuthal component of the 4-velocity. Therefore, the corresponding
non-local coordinate transformation (\ref{nd-int-1})-(\ref{nd-int-2})
produces a modification of the coordinate time $r^{0}(s)$ taking into
account also the contribution of the azimuthal velocity.

7) Also for the diagonalizing NLPT a teleparallel realization can be given.
This follows by identifying now the space-time $(\mathbf{Q}^{4},g)$ with the
Minkowski space-time. The solution for the Jacobian of a such a
transformation is obtained from Eqs.(\ref{xxx-1})-(\ref{xxx-3}) by setting $%
S_{\mu }\left( r\right) =1$ identically. This means that it is always
possible to transform a non-diagonal metric tensor into the Minkowski one by
means of the inverse diagonalizing NLPT transformation.

8) Finally, an interesting comparison is possible with the so-called
Newman-Janis algorithm \cite{NJ1,NJ2,NJ3}. As is well known (see also
related discussion in Part 1) this algorithm can be used to diagonalize
non-diagonal metric tensors and is frequently used in the literature for the
purpose of investigating a variety of standard or non-standard GR black-hole
solutions \cite{bambi,bambi2}.\ Its basic feature is that adopting a complex
coordinate transformation, a feature which effectively inhibits its physical
interpretation and puts in doubt its very validity. In contrast, within the
present NLPT approach, the physical consistency of the transformation
approach is preserved. Hence, the present conclusions seem particularly
rewarding. Indeed, based on the NLPT-approach indicated above, the
difficulties and physical limitations of the complex Newman-Janis algorithm
are effectively avoided by adopting the NLPT-theory. This is of paramount
importance for theoretical and astrophysical applications, such as the
physics around rotating black holes and gravitational waves.

\section{6 - Conclusions}

In this paper the problem has been posed of extending the class of local
point transformations (LPT) on which the general covariance principle (GCP)
lying at the bases of General Relativity (GR) is based. Such transformations
in the customary formulation of GR map in each other different reference
frames, i.e., coordinate systems. However, theoretical motivations suggest
the extension of the traditional concept of reference frame adopted
previously in GR based on the identification of an extended class of point
transformations. These have been constructed relying on a number of physical
requirements (\textit{Requirements} \textit{\#1-\#3}), prescribing in
particular the functional form of the corresponding Jacobians, and referred
to as non-local point transformations (NLPT). While extending the class of
local point transformations (LPT) on which both the differential geometry
and the GCP rely, NLPT permit one to map intrinsically physically-different
space-times, i.e., characterized by different metric and curvature Riemann
tensors. Two characteristic features of these transformations emerge. The
first one is their non-locality, which appears both in their Lagrangian and
Eulerian forms. This is due to a non-local linear dependence with respect to
the transformed $4-$velocity. The second one lies in their Jacobians. In
difference with the case of LPT, the latter by construction cannot be
identified with gradient operators. Nevertheless, since the same Jacobians
remain velocity-independent, tensor transformation laws can be still
determined, which are based on the transformation properties holding for the
infinitesimal displacements and the corresponding $4-$velocities. In
addition, a physical interpretation of NLPT has been pointed out which is
based on an ideal (Gedanken) experiment.

Two different applications of the theory have been proposed, which concern
the mapping between diagonal metric tensors and the diagonalization of
non-diagonal metric tensors. Both these problems cannot be approached in the
framework of customary LPT, while their solution becomes straightforward and
physically-consistent when the theory of NLPT developed here is invoked.

These features, in our view, suggest the theory presented here as an
extremely promising and innovative research topic, which might eventually
give rise to a novel scientific mainstream in GR. The theory developed here
is in fact susceptible of a plethora of potential applications, besides its
natural framework, i.e., GR. In particular, general NLPT-theory provides the
theoretical basis for important possible subsequent developments ranging
from classical relativistic mechanics and electrodynamics \cite%
{EPJ1,EPJ2,EPJ3,EPJ4,EPJ5,EPJ6,EPJ7}, quantum theory of extended particle
dynamics \cite{EPJ8}, relativistic kinetic theory \cite{EPJ5}, to cosmology
as well as relativistic quantum mechanics and quantum gravity.

\bigskip

\section{Acknowledgments}

Work developed within the research projects of the Czech Science Foundation
GA\v{C}R grant No. 14-07753P (C.C.)\ and Albert Einstein Center for
Gravitation and Astrophysics, Czech Science Foundation No. 14-37086G (M.T.).

\section{Appendix - Eulerian and Lagrangian forms of tensor fields}

From the definition of the Riemannian distance $ds$ (see Eq.(\ref{Riemann
distance})) it follows that%
\begin{equation}
ds=g_{\mu \nu }(r)\frac{dr^{\mu }(s)}{ds}dr^{\nu }(s),  \label{ape-1}
\end{equation}%
or equivalently%
\begin{equation}
ds=g_{\mu \nu }^{\prime }(r^{\prime })\frac{dr^{\prime \mu }(s)}{ds}%
dr^{\prime \nu }(s).  \label{ape-2}
\end{equation}%
Hence, integrating and letting $s_{o}=0$ one obtains:%
\begin{equation}
s-s_{o}\equiv s=\int_{r^{\nu }(s_{o})}^{r^{\nu }(s)}g_{\mu \nu }(r)\frac{%
dr^{\mu }(s^{\prime })}{ds^{\prime }}dr^{\nu },  \label{A-1}
\end{equation}%
where the integration variable $r^{\nu }$ belongs to the $4-$dimensional
subset of $(\mathbf{Q}^{4},g)$, the set having boundaries $r^{\nu }(s_{o})$
and $r^{\nu }(s)$. Notice furthermore that in the integrand on the rhs of
the previous equation the variable $s^{\prime }$ is to be considered as
dependent from the integration variable, i.e., of the form $r^{\nu }\equiv
r^{\nu }(s^{\prime })$. Indeed, from Eq.(\ref{A-1}) it follows manifestly
also that%
\begin{equation}
s^{\prime }=\int_{r^{\nu }(s_{o})}^{r^{\nu }(s^{\prime })}g_{\mu \nu }(r)%
\frac{dr^{\mu }(\overline{s})}{d\overline{s}}dr^{\nu }.  \label{A-2}
\end{equation}%
Hence, Eq.(\ref{A-1}) implies necessarily that%
\begin{equation}
s=s(r),  \label{A-3}
\end{equation}%
where $r\equiv r^{\nu }\equiv \left\{ r^{\nu }(s)\right\} $ and similarly
form Eq.(\ref{A-2}) it follows that $s^{\prime }=s^{\prime }(r(s^{\prime }))$%
.

Let us now consider an arbitrary tensor field $A_{\mu \nu }$ - for example
to be identified with the metric tensor $g_{\mu \nu }$ as in Section 3 -
which when expressed in Lagrangian form is assumed to take the form $A_{\mu
\nu }(r,s)$. Here $r$ denotes $r=r(s)\equiv \left\{ r^{\nu }(s)\right\} $,
while for all $s\in I\equiv
\mathbb{R}
$, $s$ is the proper-time which is associated with the time-like world line.


\begin{thebibliography}{99}
\bibitem{Part 1} M. Tessarotto, C. Cremaschini, \textit{Theory of non-local
point transformations - Part 1: \ Representation of Teleparallel Gravity,}
Eur. Phys. J. Plus, submitted (2015).

\bibitem{Einstein1915} A. Einstein, \textit{Die Feldgleichungen der
Gravitation, }Sitzungsber, Preuss. Akad. Wiss. (Berlin), 844 (1915).

\bibitem{ein-1907} A. Einstein, \textit{Relativit\"{a}tsprinzip und die aus
demselben gezogenen Folgerungen} (\textit{On the Relativity Principle and
the Conclusions Drawn from It}). Jahrbuch der Radioaktivit\"{a}t \textbf{4},
411 (1907).

\bibitem{ein-1911} A. Einstein, \textit{Einfluss der Schwerkraft auf die
Ausbreitung des Lichtes} (\textit{On the Influence of Gravitation on the
Propagation of Light}), Annalen der Physik \textbf{35}, 898 (1911).

\bibitem{ein28} A. Einstein, \textit{Riemann-Geometrie mit Aufrechterhaltung
des Begriffes des Fernparallelismus.} Preussische Akademie der
Wissenschaften, Phys.-math. Klasse, Sitzungsberichte 217 (1928).

\bibitem{Einst} A. Einstein, \textit{The Meaning of Relativity,} Princeton
University Press (1945).

\bibitem{Cremaschini2015} C. Cremaschini, M. Tessarotto,\textit{\ }Eur.
Phys. J. Plus \textbf{130}, 123 (2015).

\bibitem{Landau} L.D .Landau, V. Lifschitz, \textit{The Classical Theory of
Fields,} Vol.2 (Addison-Wesley, N.Y., 1957).

\bibitem{Wheeler} J.A.. Wheeler, C. Misner and K.S., Thorne, \textit{%
Gravitation}. W.H. Freeman \& Co (1973).

\bibitem{wald} R.M.Wald, \textit{General Relativity}. University of Chicago
Press, 1st edition (1984).

\bibitem{Synge} J.L. Synge and A. Schild, \textit{Tensor Calculus,\ }Dover
Publications 1978 edition. pp. 6--108 (1949).

\bibitem{NJ3} S.P. Drake and P. Szekeres, Gen. Relativ. Gravit. \textbf{32},
445 (2000).

\bibitem{bambi} C. Bambi and L. Modesto, Phys. Lett. B \textbf{721}, 329
(2013).

\bibitem{bambi2} B. Toshmatov, B. Ahmedov, A. Abdujabbarov and Z. Stuchl%
\'{\i}k, Phys. Rev. D \textbf{89}, 104017 (2014).

\bibitem{NJ1} E.T. Newman and A.I. Janis, J. Math. Phys. \textbf{6}, 915
(1965).

\bibitem{NJ2} E.T. Newman, E. Couch, K. Chinnapared, A. Exton, A. Prakash
and R. Torrence, J. Math. Phys. \textbf{6}, 918 (1965).

\bibitem{EPJ1} C. Cremaschini, M. Tessarotto, Eur. Phys. J. Plus \textbf{126}%
, 42 (2011).

\bibitem{EPJ2} C. Cremaschini, M. Tessarotto, Eur. Phys. J. Plus \textbf{126}%
, 63 (2011).

\bibitem{EPJ3} C. Cremaschini, M. Tessarotto, Eur. Phys. J. Plus \textbf{127}%
, 4 (2012).

\bibitem{EPJ4} C. Cremaschini, M. Tessarotto, Eur. Phys. J. Plus \textbf{127}%
, 103\textbf{\ }(2012).

\bibitem{EPJ5} C. Cremaschini, M. Tessarotto, Phys. Rev. E \textbf{87},
032107 (2013).

\bibitem{EPJ6} C. Cremaschini, M. Tessarotto, Int. J. Mod. Phys. A \textbf{28%
}, 1350086 (2013).

\bibitem{EPJ7} C. Cremaschini, M. Tessarotto, Eur. Phys. J. Plus \textbf{129}%
, 247 (2014).

\bibitem{EPJ8} C. Cremaschini, M. Tessarotto,\textit{\ }Eur. Phys. J. Plus
\textbf{130}, 166 (2015).
\end{thebibliography}
\end{document}